\documentclass[preprint]{elsarticle}

\usepackage{array}
\usepackage{dcolumn}% Align table columns on decimal point
\usepackage{url}
\usepackage{breakurl}

\usepackage{bm}% bold math
\usepackage[colorlinks, pdfborder={0 0 0}, plainpages=false]{hyperref}
\usepackage{graphicx}
\usepackage{xspace}
\usepackage[usenames,dvipsnames]{color}
\usepackage{amssymb}
\usepackage[normalem]{ulem} %for \sout
\usepackage{letltxmacro}
\usepackage{amsfonts}
\usepackage{amsmath}
\usepackage{amssymb}
\usepackage{bm}
\usepackage{dcolumn}
\usepackage{epsfig}
\usepackage{graphics}
\usepackage[latin1]{inputenc}
\usepackage{latexsym}
\usepackage{rotating}
\usepackage{hyperref}
\usepackage[usenames]{color}
\usepackage{float}
\usepackage{xspace} 
\usepackage{mathrsfs}
\usepackage{subfigure}
\usepackage{multirow}
\usepackage{booktabs}
\usepackage{soul}

\usepackage{colortbl}
\usepackage{etoolbox}
\usepackage[table]{xcolor}

\DeclareMathOperator*{\argmin}{arg\,min}

\def\lesssim{\mathrel{\hbox{\rlap{\hbox{\lower4pt\hbox{$\sim$}}}\hbox{$<$}}}}
\def\gtrsim{\mathrel{\hbox{\rlap{\hbox{\lower4pt\hbox{$\sim$}}}\hbox{$>$}}}}
\def\alt{\mathrel{\hbox{\rlap{\hbox{\lower4pt\hbox{$\sim$}}}\hbox{$<$}}}}
\def\agt{\mathrel{\hbox{\rlap{\hbox{\lower4pt\hbox{$\sim$}}}\hbox{$>$}}}}

\usepackage{mathtools}

\def\gta{\ifmmode {\mathbin{\lower 3pt\hbox   %> or of order
    {$\,\rlap{\raise 5pt\hbox{$\char'076$}}\mathchar"7218\,$}}}
    \else {${\mathbin{\lower 3pt\hbox
    {$\rlap{\raise 5pt\hbox{$\char'076$}}\mathchar"7218\,$}}}
    $}\fi}
\def\lta{\ifmmode {\,\mathbin{\lower 3pt\hbox   %< or of order
    {$\,\rlap{\raise 5pt\hbox{$\char'074$}}\mathchar"7218\,$}}}
    \else {${\mathbin{\lower 3pt\hbox
    {$\rlap{\raise 5pt\hbox{$\char'074$}}\mathchar"7218\,$}}}
    $}\fi}

\newcommand{\msun}{{\rm M}_{\odot}}
\newcommand{\beq}{\begin{equation}}
\newcommand{\eeq}{\end{equation}}
\newcommand{\bea}{\begin{eqnarray}}
\newcommand{\eea}{\end{eqnarray}}

\definecolor{darkperiwinkle}{RGB}{102, 102, 128}

\definecolor{light-gray}{gray}{0.9}

\newcommand{\RNum}[1]{\uppercase\expandafter{\romannumeral #1\relax}}

\usepackage{lineno,hyperref}
\modulolinenumbers[5]

\journal{Journal of \LaTeX\ Templates}

\begin{document}

\begin{frontmatter}

\title{Gravitational Wave Denoising\\ of Binary Black Hole Mergers\\ with Deep Learning}


\author[NCSA,PNCSA]{Wei Wei}
\author[NCSA,ANCSA]{E. A. Huerta}

\address[NCSA]{NCSA, University of Illinois at Urbana-Champaign, Urbana, Illinois 61801, USA}
\address[PNCSA]{Department of Physics, University of Illinois at Urbana-Champaign, Urbana, Illinois 61801, USA}
\address[ANCSA]{Department of Astronomy, University of Illinois at Urbana-Champaign, Urbana, Illinois 61801, USA}

\begin{abstract}
\noindent Gravitational wave detection requires an in-depth understanding of the physical properties of gravitational wave signals, and the noise from which they are extracted. Understanding the statistical properties of noise is a complex endeavor, particularly in realistic detection scenarios. In this article we demonstrate that deep learning can handle the non-Gaussian and non-stationary nature of gravitational wave data, and showcase its application to denoise the gravitational wave signals generated by the binary black hole mergers GW150914, GW170104, GW170608 and GW170814 from advanced LIGO noise. To exhibit the accuracy of this methodology, we compute the overlap between the time-series signals produced by our denoising algorithm, and the numerical relativity templates that are expected to describe these gravitational wave sources, finding overlaps \({\cal{O}}\gtrsim0.99\). We also show that our deep learning algorithm is capable of removing noise anomalies from numerical relativity signals that we inject in real advanced LIGO data. We discuss the implications of these results for the characterization of gravitational wave signals.
\end{abstract}


\end{frontmatter}

%\linenumbers

\section{Introduction}
\label{intro}

Gravitational wave (GW) observations of binary black hole (BBH) mergers 
with the LIGO~\cite{DII:2016,LSC:2015} and Virgo~\cite{Virgo:2015} 
detectors is now a common occurrence~\cite{DI:2016,secondBBH:2016,thirddetection,fourth:2017,GW170608,CatalogBBH:2018}. Extracting these time-series 
signals from non-Gaussian and non-stationary noise requires a firm understanding 
of the astrophysical properties of GWs, which is customarily obtained through 
numerical relativity (NR) simulations~\cite{Chu:2016CQG,Mroue:2013,Kumar:2016dhh,NRI:2016,NRGW1509142,Lange:2017PhRvD,Mroue:2013,RITcatalog:2017,GaTechcatalog:2016,jade1:2016}. 
On the other hand, characterizing noise in GW detectors is a daunting task. Its 
non-Gaussian and non-stationary nature, combined with the fact that GW facilities 
undergo frequent commissioning to further enhance their sensitivity, presents a 
formidable challenge to design robust models that can accurately capture its statistical properties~\cite{shengeorge:PhysRevD97,Gspy:2017,jade:2015CQGra,cava:2018C}.
Nonetheless, this work is critical to identify and excise poor quality data segments, and 
noise anomalies that contaminate GW signals. Once this is done, GW detection pipelines can 
provide robust estimates for the astrophysical parameters of GW sources, and their significance.

While noise anomaly removal is customarily done in off-line GW searches, low-latency detection pipelines 
also require data quality information to identify and remove, in real-time, noise anomalies that 
may prevent the detection of GW events, and to accurately determine their nature, which is of 
central importance for Multi-Messenger Atrophysics  searches~\cite{bnsdet:2017,mma:2017arXiv,2017Sci...358.1556C,kiloGW170817:2017,grb:2017ApJ}. To complement this ongoing effort,  in this article 
we present deep learning algorithms that 
are trained with raw advanced LIGO noise to identify GW signals in realistic detection scenarios, and which upon 
removing the imprints of noise, produce denoised time-series signals that resemble NR waveforms. 
Given that BBHs represent the most abundant source of GW sources thus far~\cite{CatalogBBH:2018}, 
the analysis we present herein focuses on GWs produced by BBH mergers. When we apply these algorithms to denoise several BBH waveforms that have been detected by the advanced LIGO and Virgo detectors, we find that the output time-series data of our denoising algorithm reproduces with excellent accuracy the NR templates that optimally describe these GW sources. Furthermore, we also demonstrate that when we contaminate NR templates with simulated noise anomalies, following~\cite{jade:2015CQGra,jade1:2016}, and inject these signals in real advanced LIGO noise, our denoising algorithm can tell apart between true GW waveform signals and glitches. 

Denoising gravitational wave signals in low-latency may be useful for a variety of tasks. First and foremost, it may be readily applied to remove noise anomalies that contaminate or obscure true gravitational wave signals. Once this is done, denoised GW signals may be used in conjunction with existing algorithms to assess the significance of new detections. Furthermore, denoised time-series waveforms may be used to compute fast time-domain overlap calculations with machine learning based waveform generators~\cite{Varma:2019csw,varma:smhprd} to constrain the astrophysical parameters of the source, which in turn may be used to inform the construction of physical priors for parameter estimation analyses, or to explore whether it is necessary to produce numerical relativity waveforms to accurately describe an event that is beyond the scope of existing semi-analytical waveform models.

This work aims to accelerate the convergence of novel signal-processing algorithms with GW astrophysics. Recent accomplishments of this program include the demonstration of 
deep learning for the detection and characterization of GW signals in simulated and real LIGO noise~\cite{geodf:2017a,geodf:2017b,geoNIPS:2017}, the detection and characterization of higher-order waveform signals from eccentric 
BBH mergers~\cite{Adam:2018arXiv}, among many recent applications of machine and deep learning for signal detection and source modeling~\cite{shengeorge:PhysRevD97,AlvinC:2018,Fan:2018,Gonza:2018,2018GN,Fuji:2018,LiYu:2017,hshen:2017,Nakano:2018}. 

In the specific context of signal denoising, recent efforts have focused on the use of recurrent neural networks (RNNs) ~\cite{279181,Hochreiter:1997:LSM:1246443.1246450} combined with auto-encoders, dictionary learning and principal component analysis to denoise burst-like GWs, i.e., short duration (\({\cal{O}}(10^{-1}\textrm{second})\)) signals with large signal-to-noise ratios (SNRs)~\cite{lorens:2018,tvm:2018PhRvDT,hshen:2017}. While recurrent auto-encoders have been proven to outperform principal component analysis and dictionary learning both in signal reconstruction accuracy and computational efficiency~\cite{hshen:2017}, it has thus far been difficult to extend these algorithms to denoise \({\cal{O}}(\textrm{second-long}\)) GW signals. 

Motivated by the fact that ground-based GW detectors continue to enhance their sensitivity, thereby increasing the time-window during which GW signals can be observed, we have exhaustively explored the use of different types of neural network models, and have found that convolutional neural networks (CNNs)~\cite{726791,krizhevsky2012imagenet} are better suited to denoise BBH GW signals whose length is \(\sim10\textrm{x}\) longer, and whose SNR are significantly lower, than what existing state-of-the-art algorithms can handle. 

We have tested our new CNN-based algorithm in a variety of scenarios, including the denoising of GW signals in simulated Gaussian noise, and the denoising of true BBH GW signals in realistic detection scenarios. These results represent the first application of deep learning to remove 
noise contamination from true BBH GW events that span a wide range of masses and SNRs. By computing the 
overlap between the output of our denoising algorithms with the optimal NR waveforms that describe these 
signals, we furnish evidence for the robustness and accuracy of this approach. 

This article is organized as follows. We describe the properties of our deep learning denoising algorithm, and the datasets used to train it in Section~\ref{drl}. Results for the denoising of GW signals in simulated and real LIGO noise are presented in Section~\ref{sec:results}. We summarize out findings and future directions of work in Section~\ref{end}.

%%%%%%%%%%%%%%%%%%%%%%%%%%%%%%%%%%%%%%%%%%%%%
%%%%%%%%%%%%%%%%%%%%%%%%%%%%%%%%%%%%%%%%%%%%%

\section{Methods}
\label{drl}

In this section we provide a succinct overview of the 
mathematical and statistical foundation of the signal processing algorithms we have 
utilized for GW denoising. Thereafter, we 
describe the architecture and key features of our neural network models, and 
the datasets we have used to train and test them to denoise GWs, both in the context of 
simulated Gaussian noise and real LIGO noise. 

\subsection{Statistical foundations of Deep Learning Denoisers}
\label{stats}

Within the framework of statistical learning, a GW signal $X$ can be modeled as a random process, indexed by real time $t$. Since we use modeled one-second GWs sampled at 8192 Hz, we treat GWs as random vectors of size 8192. We standardize our datasets by normalizing their peak amplitude so that the sample space $\Omega$ can be set to $[-1,1]^{8192}$. 

We assume that GWs follow some unknown but fixed joint probability distribution, with the probability density function (pdf) $f_X(x)$. The GW signal contaminated by noise is denoted by $Y$, and follows some unknown but fixed  distribution $f_{Y|X}(y|x)$ when conditioned on some clean signal $x$. Under these conventions, the goal of denoising is to find a function $h(\cdot)$ that minimizes the expectation value of the mean square error (MSE) of the recovered signal, namely

\begin{align}
    L(h) = \int\left[\int\Vert h(y)-x\Vert^2 f_{Y|X}(y|x)dy\right]f_X(x)dx\,,
\end{align}

\noindent where $h(\cdot)$ is the denoising function. In most cases, we only know the empirical distribution $\hat{f}_{X}(x)$ of $X$ and $\hat{f}_{Y|X}(y|x)$ of $Y$, which are determined by the empirical data. So the quantity we can directly minimize is

\begin{align}
    \hat{L}(h) = \int\left[\int\Vert h(y)-x\Vert^2 \hat{f}_{Y|X}(y|x)dy\right]\hat{f}_{X}(x)dx\,.
\end{align}

\noindent In practice, if the choice of $h(\cdot)$ is arbitrary, then finding an optimal solution is computationally unfeasible. Therefore, we often restrict the searching space to a class of parameterized functions, $h_{\mathbf{w}}(\cdot)$, where $\mathbf{w}$ is a vector of parameters. In this case, the optimization problem can be posed as 

\begin{align}\label{eq:min}
    \mathbf{w}^{*}=\argmin_{\mathbf{w}} \hat{L}(h_{\mathbf{w}})\,.
\end{align}

\noindent The choice of the parameterized function class is critical to the success of any statistical learning algorithm. In recent years, a deep-layered structure of functions has received much attention ~\cite{lecun2015deep,Goodfellow-et-al-2016},

\begin{align}
    h_{\mathbf{w}}(x)=h_{\mathbf{w}_n}(h_{\mathbf{w}_{n-1}}(\cdots h_{\mathbf{w}_1}(x))),
\end{align}

\noindent where \(n\) is the number of layers or the depth. Usually, we choose, \(h_{\mathbf{w}_i}(\mathbf{x})=g(\mathbf{w}_i\mathbf{x})\), where \(\mathbf{w}_i\) is a matrix, \(\mathbf{x}\) is an input vector, and \(g(\cdot)\) is a fixed non-linear function, e.g., $\max\{\cdot,0\}$ (also known as \texttt{ReLU}), \(\tanh(\cdot)\), etc, that is applied element-wise. This function class and its extensions, also dubbed neural networks, combined with simple first-order optimization algorithms such as stochastic gradient descent (SGD), and improved computing hardware, has lead to disruptive applications of deep learning~\cite{lecun2015deep,Goodfellow-et-al-2016}.

\subsection{Neural network architecture}

Empirically, it has been shown that a particular function class or network structure called \texttt{WaveNet}~\cite{oord2016wavenet} can be used to produce raw audio waveforms that mimic human speech with high fidelity. In view of this realization, we have explored this architecture as a starting point to design a neural network model to denoise GWs. Since we are using \texttt{WaveNet} for denoising purposes, instead of waveform generation, we have removed the causal structure of the network. The causal structure of \texttt{WaveNet} is modeled with a convolutional layer~\cite{krizhevsky2012imagenet} with kernel size 2, and by shifting the output of a normal convolution by a few time steps. However, in this paper we adopt convolutional layers with kernel size 3, so that when denoising the waveform at a certain time step, we take into account information from past and future time steps. We also dilate the convolutional layers to get an exponential increase in the size of the receptive field~\cite{oord2016wavenet}. This is necessary to capture long-range correlations, as well as to increase computational efficiency.  By construction, \texttt{WaveNet} utilizes deep residual learning, which is specifically tailored to train deeper neural network models~\cite{he2016deep}. The structure of \texttt{WaveNet} is described in detail in~\cite{oord2016wavenet}, and we provide a schematic representation of its architecture in Figure~\ref{fig:net}. To demonstrate the robustness of the performance of \texttt{WaveNet} for denoising, we consider two models with different sets of hyper-parameters, which we describe below. 

\subsubsection{Model \RNum{1}}
\begin{figure}
\centerline{
\includegraphics[width=0.8\textwidth]{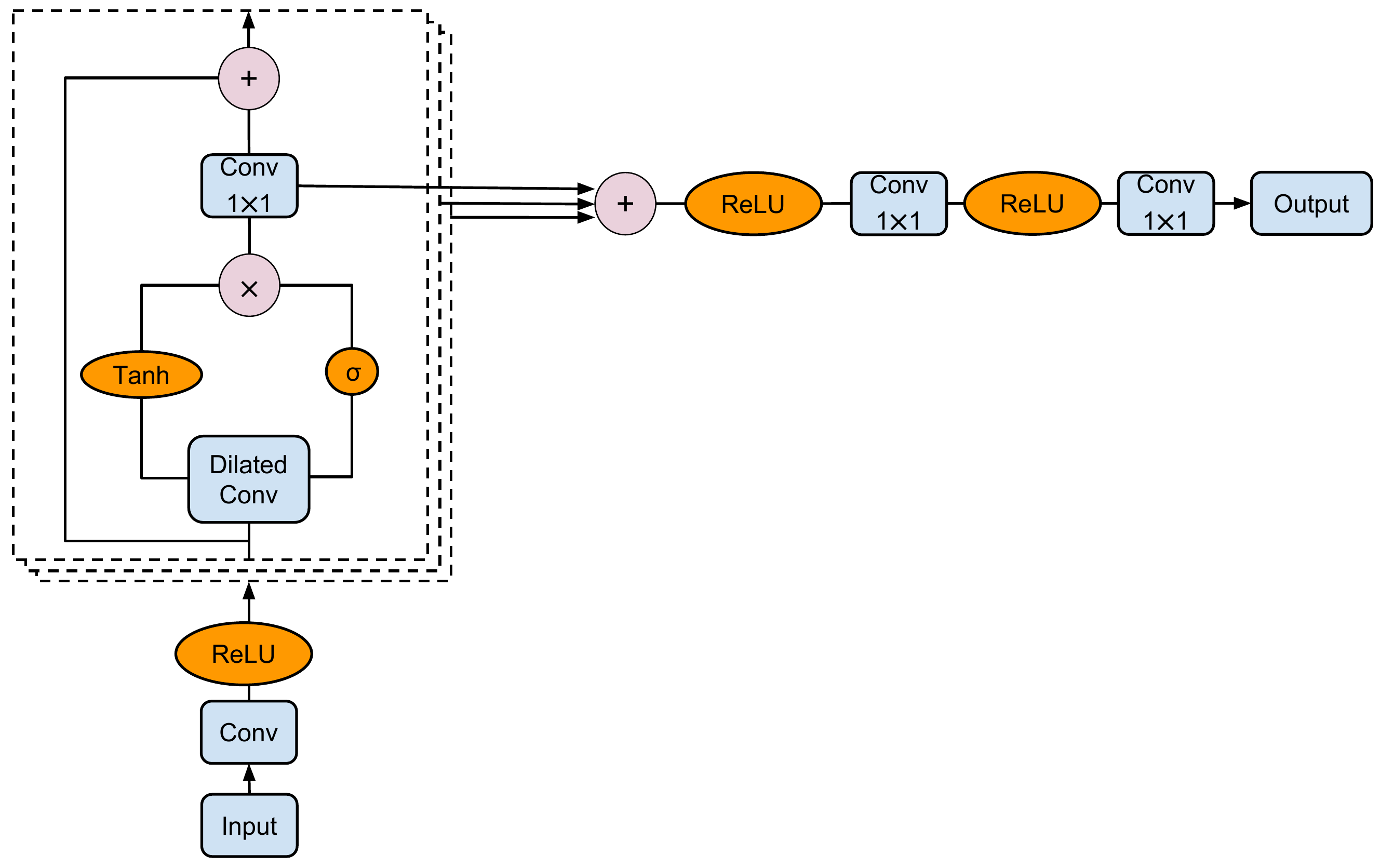}\hspace{-5.5cm}
\raisebox{0.5cm}{
\includegraphics[width=0.5\textwidth]{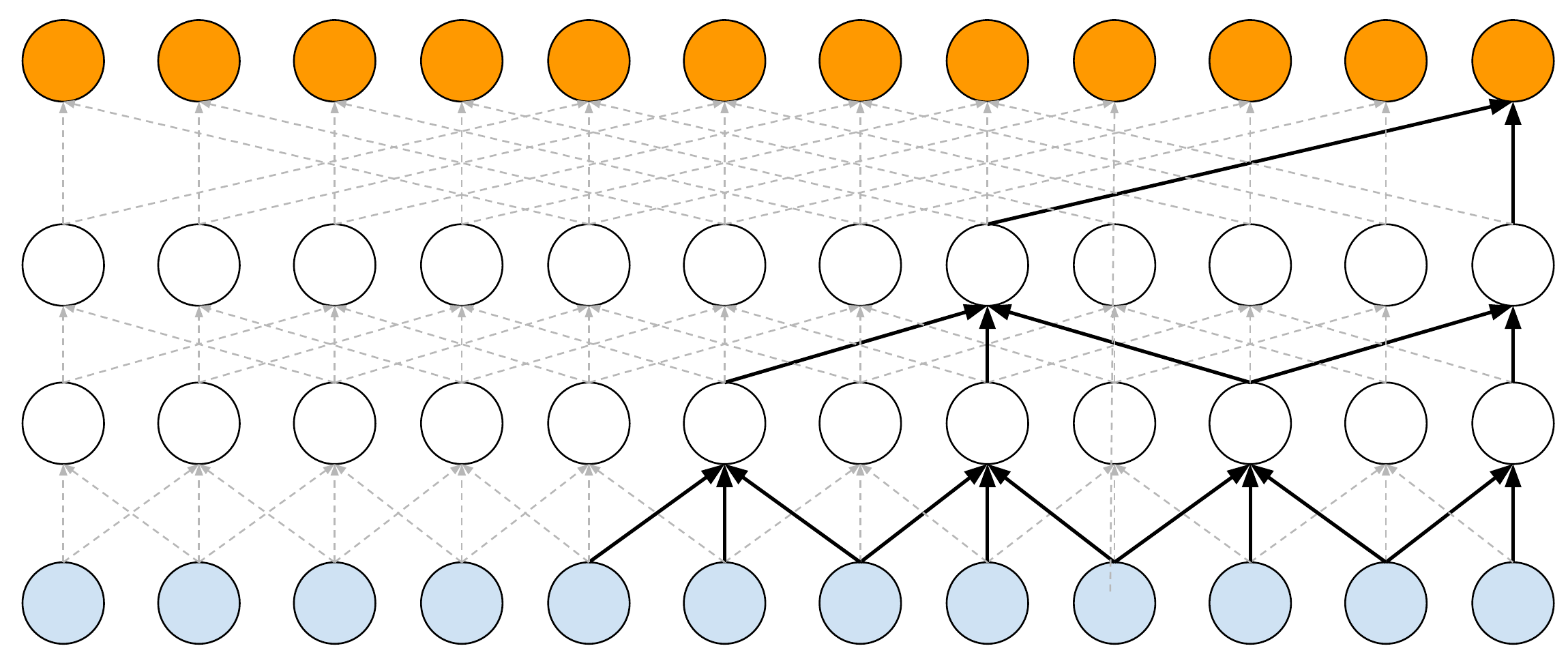}
}
} 
\caption{Left panel: architecture of our \texttt{WaveNet} denoising algorithm.  The input and output are tensors of shape $\textrm{batch\_size}\times1\times8192$. As shown in the figure, we have stacked together several layers (boxed regions), and that for each of these layers the dilation size is a power of $2$, as shown in the bottom-right panel. Notice that a skip connection also branches out from each layer to the final layers to improve the back-propagation during the training. Bottom-right panel: dilated convolutional layers. The kernel size is 3, and zeroes are padded to both sides of the input (not shown in the figure) to make sure that the output and input have the same size. The dilation is set to be the power of 2. The figure shows convolution layers with dilations 1, 2 and 4, from the second lowest to the uppermost layer.}
\label{fig:net}
\end{figure}

For Model \RNum{1}, the dilated convolutional layers have dilations $2^0,2^1,2^2,2^3,...2^{10}$. These 11 layers are stacked as a block, which is repeated ten times. The non-dilated convolutional layers in the repeating blocks (Conv $1\times 1$ in the boxes of Figure~\ref{fig:net}) each use a kernel size of 1.  Furthermore, the numbers of input and output channels are 128 for both dilated and non-dilated convolutional layers in the repeating blocks. The penultimate convolutional layer (Conv $1\times 1$ in the middle of Figure~\ref{fig:net}) has 128 input channels, and 64 output channels with kernel size 1. The last convolutional layer (rightmost Conv $1\times 1$ in Figure~\ref{fig:net}) has 64 input channels, 1 output channel and kernel size 1.

\subsubsection{Model \RNum{2}}
\label{m2}

 We use a similar structure for Model \RNum{2}, except for the dilated convolutional layers, which now have dilations $2^0,2^1,2^2,2^3,...2^{11}$. These twelve layers are stacked as a block and repeated six times. Additionally, the number of input and output channels in the convolutional layers are increased from 128 to 256 and 64 to 128, respectively. 
 
 We have considered these two models to assess their robustness and accuracy to remove noise contamination from GW signals. When applied in the context of simulated Gaussian noise or real LIGO noise, we have found that the denoised time-series signals obtained from either model are identical. These findings, shown below, demonstrate the robustness of \texttt{WaveNet} to denoise GW signals in realistic detection scenarios. 
 
\subsection{Data Curation}
\label{sec:curation}

We trained our denoising algorithms using a catalog of GW signals that describe the inspiral, merger and ringdown of non-spinning BBH mergers. We produced these waveforms using the NR surrogate waveform family~\cite{blackman:2015}. Each waveform is produced at sample rate of 8192Hz, and we consider the last second of evolution of BBHs with component masses \(m_{\{1,2\}}\in[5\msun,\,75\msun]\), and mass-ratio \(q \leq 10\). The training dataset (9861 templates) samples this parameter space in steps of \(\msun\), while the testing dataset (2459 templates) comprises intermediate masses in this range. 

We start the training stage by computing the Power Spectral Density (PSDs) of the noise, which is used to whiten both the templates and the noise. Thereafter, we rescale the amplitude of the GW signals, and the standard deviation of the noise to create scenarios that describe GWs over a wide range of SNRs. We then add the rescaled templates and noise together, and normalize the standard deviation of data that contains both signals and noise. The simulated noisy signals will be used as input, and the corresponding clean signals will be used as targets during the training process. The actual output of our denoising algorithm is an unwhitened waveform signal. In our results, we also present the corresponding whitened signals, both denoised and true GW signals, to clearly show what portion of the denoised time-series signal is actually detectable by the advanced LIGO detectors. To further enhance the robustness of the network, we have incorporated time invariance, which basically ensures that the neural network can correctly identify and denoise signals, irrespective of their location in the data stream. 

\noindent \textit{Signals embedded in simulated Gaussian noise} We followed the previous methodology using simulated Gaussian noise, and setting LIGO's Zero Detuned High Power (ZDHP) configuration as the target PSD~\cite{ZDHP:2018}.

\noindent \textit{Signals embedded in real LIGO noise} Following well established methods to measure a noise PSD estimate~\cite{2016CQGra..33u5004U}, and to encapsulate the actual sensitivity of the advanced LIGO detectors at the time true BBH GW signals were observed, we use between 512 seconds and 4096 seconds of open source advanced LIGO data around the GWs we want to denoise. It is worth noting that this PSD estimate only needs to be regenerated when there are significant changes in the detector's noise PSD, as described in~\cite{2016CQGra..33u5004U,Littenberg:2014oda}. In practice this means that it suffices to compute a noise PSD estimate every 4096 seconds~\cite{2016CQGra..33u5004U,Littenberg:2014oda}, and use it to denoise on the fly any new GW events that are detected within the next 4096 second interval. Doing transfer learning to continually update our neural network model with new data to capture any significant changes in the detectors' noise PSD is computationally inexpensive. One inexpensive GPU suffices to complete this task within a few minutes. It is worth mentioning that this continuous training scheme is needed, since we found that the noise PSD estimate we compute to denoise GWs in advanced LIGO's first observing run was significantly different to the noise PSDs we used to denoise GWs in advanced LIGO-Virgo's second observing run. This is expected, since the advanced LIGO detectors underwent a significant sensitivity upgrade at the end of its first observing run.

Similarly, we have followed the above description for the training and testing procedure, with the difference that we now use open source LIGO noise, available at the Gravitational Wave Open Science Center~\cite{losc}. We work with the 16384Hz LIGO noise, and downsample it as appropriate. 

The neural networks are trained on 4 NVIDIA K80 GPUs with PyTorch ~\cite{paszke2017automatic} using ADAM ~\cite{kingma2014adam} optimized method. The weight parameters are initialized randomly. The learning rate is set to $10^{-3}$ initially and reduced to $10^{-4}$ when the MSE loss plateaus.

\section{Results}
\label{sec:results}

We have tested our denoising algorithm in the context of simulated Gaussian noise, and in realistic detection scenarios using raw LIGO noise. The first set of results is presented in the following section. 

\subsection{Simulated Gaussian noise} 

Assuming that \(h\) represents the output of our denoising algorithm, \(s\) the ground truth (clean GW signal), and defining \(S_n(f)\) as LIGO's  ZDHP PSD~\cite{ZDHP:2018}, and \(\tilde{h}(f)\) as the Fourier transform of \(h(t)\), the noise-weighted inner product between \(h\) and \(s\) is given by

\beq
\left( h | s\right) = 2 \int^{f_1}_{f_0} \frac{\tilde{h}^{*}(f)\tilde{s}(f) + \tilde{h}(f)\tilde{s}^{*}(f) }{S_n(f)}\mathrm{d}f\,,
\label{inn_pro}
\eeq

\noindent with \(f_0=15\,{\rm Hz}\) and \(f_1=4096\,{\rm Hz}\).  Additionally, the normalized overlap is defined as 

\begin{align}
\label{over}
{\cal{O}}  (h,\,s)&= \underset{ t_c\, \phi_c}{\mathrm{max}}\left(\hat{h}|\hat{s}_{[t_c,\,  \phi_c]}\right)\quad{\rm with}\\
\label{n_overl}
\hat{h}&=h\,\left(h | h\right)^{-1/2}\,,
\end{align}

\begin{figure}%[h]
\centering
\includegraphics[width=0.9\linewidth]{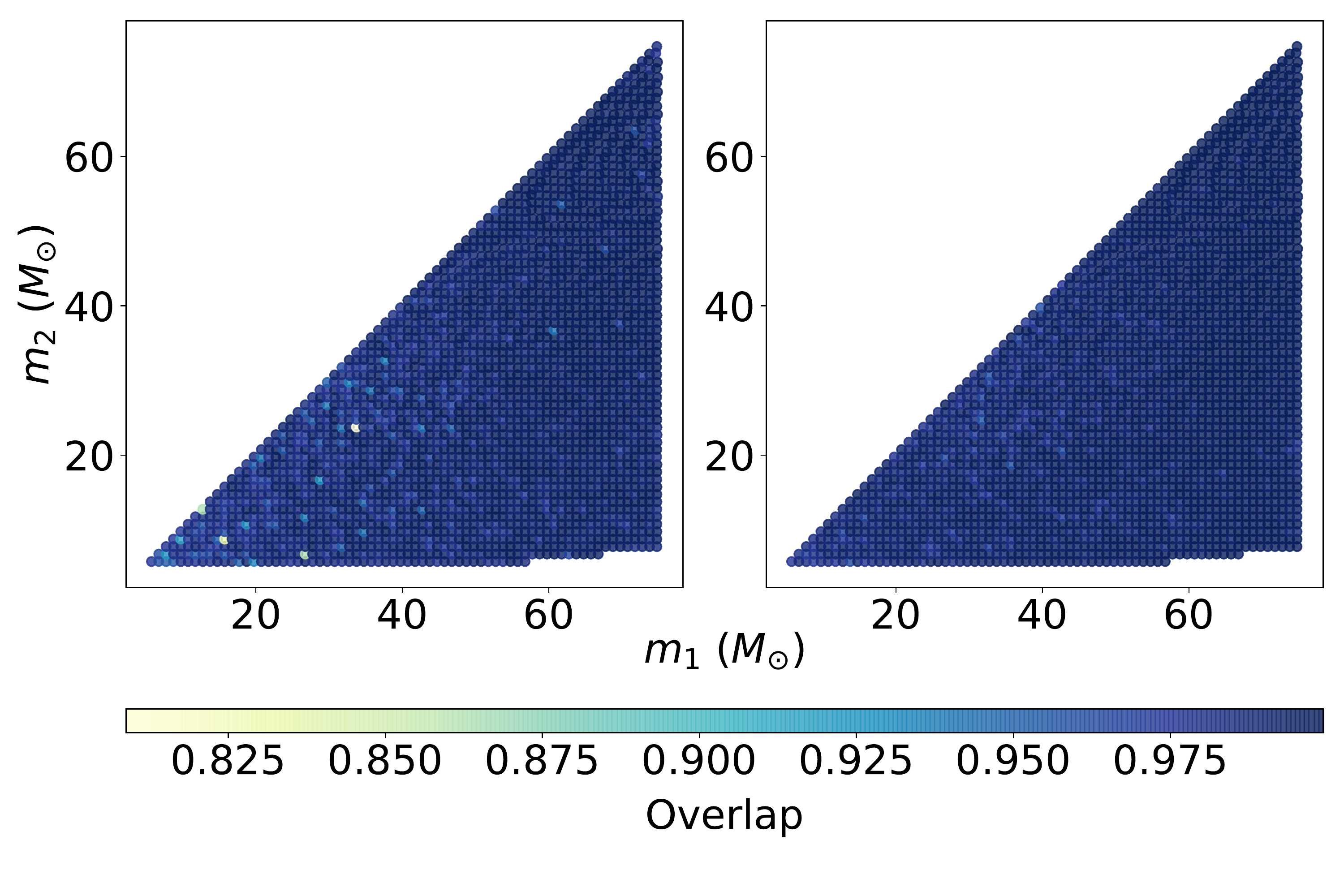}   
 \includegraphics[width=0.9\linewidth]{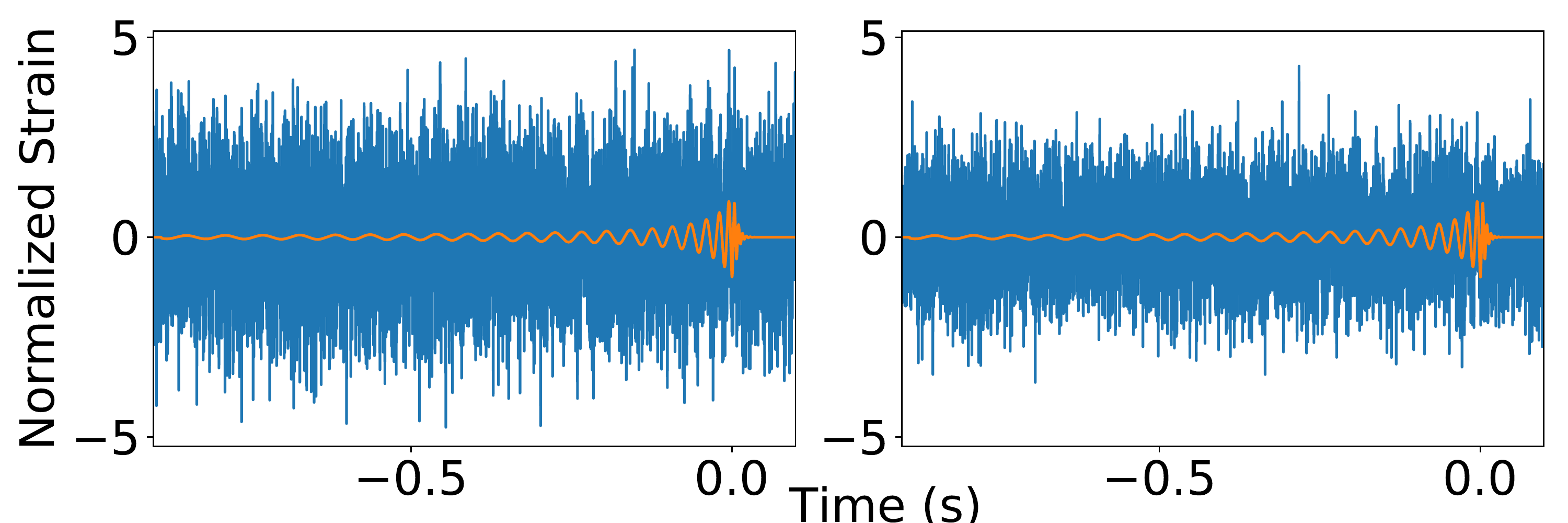} 
\caption{Top panels: normalized overlap between the output signal of our denoising algorithm (Model \RNum{1}), and the corresponding clean signals using BBH populations with matched-filtering \(\textrm{SNR}=9\) (left panel) and matched-filtering \(\textrm{SNR}=12\) (right panel). Bottom panels: sample signals from each corresponding BBH population shown in the top panels.}
\label{gaussian}
\end{figure}

\noindent where \(\hat{s}_{[t_c,\,  \phi_c]}\)  indicate that the normalized waveform \(\hat{s}\)  has been time- and phase-shifted. Under these considerations, we have used the normalized overlap to quantify the accuracy with which our denoising algorithm can reconstruct GW signals contaminated by simulated Gaussian noise.

Figure~\ref{gaussian} presents the normalized overlap between denoised signals and their clean counterparts for BBH populations with matched-filtering \(\textrm{SNR}=9\) (top left panel) and \(\textrm{SNR}=12\) (top right panel). A sample waveform embedded in noise for each BBH population is presented in the bottom panels. These results indicate that the output time-series signals of our denoising algorithm reproduce the true features of clean GW templates with overlaps \({\cal{O}}\geq0.97\) across the BBH parameter space for noisy signals with \(\textrm{SNR}\geq12\). These results were obtained with Model \RNum{1}. Results using Model \RNum{2} are presented in Figure~\ref{fig:gaussian_256}. A direct comparison between these two sets of results confirm that variants in the architecture of our neural network models produce consistent results, providing evidence for their robustness and stability when applied to denoise GW signals. 

\subsection{Real LIGO Noise} 

We have put at work Model \RNum{1} and Model \RNum{2} to denoise true BBH GW signals in real LIGO noise. We present one set of results, since the overlap between the denoised signals produced by either model, and the clean target signals, are consistent within \(1\%\).

The data selected for denoising corresponds to that in which the events are observed with the largest SNR. This choice is motivated by the results presented in Figure~\ref{gaussian}, which indicate that larger SNR values improve the waveform reconstruction.

\begin{figure*}
\centerline{
\includegraphics[width=0.25\linewidth]{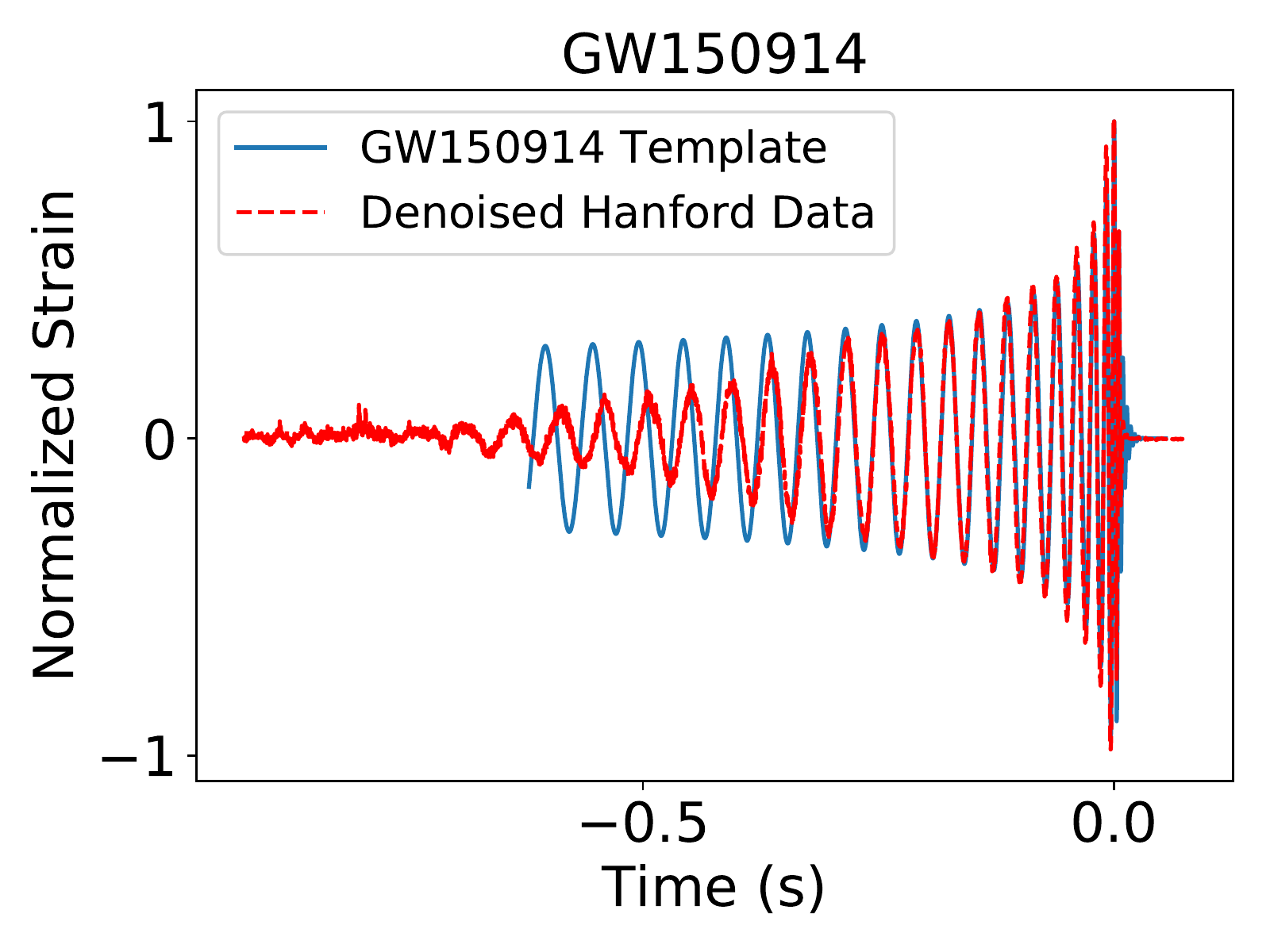}
\includegraphics[width=0.25\linewidth]{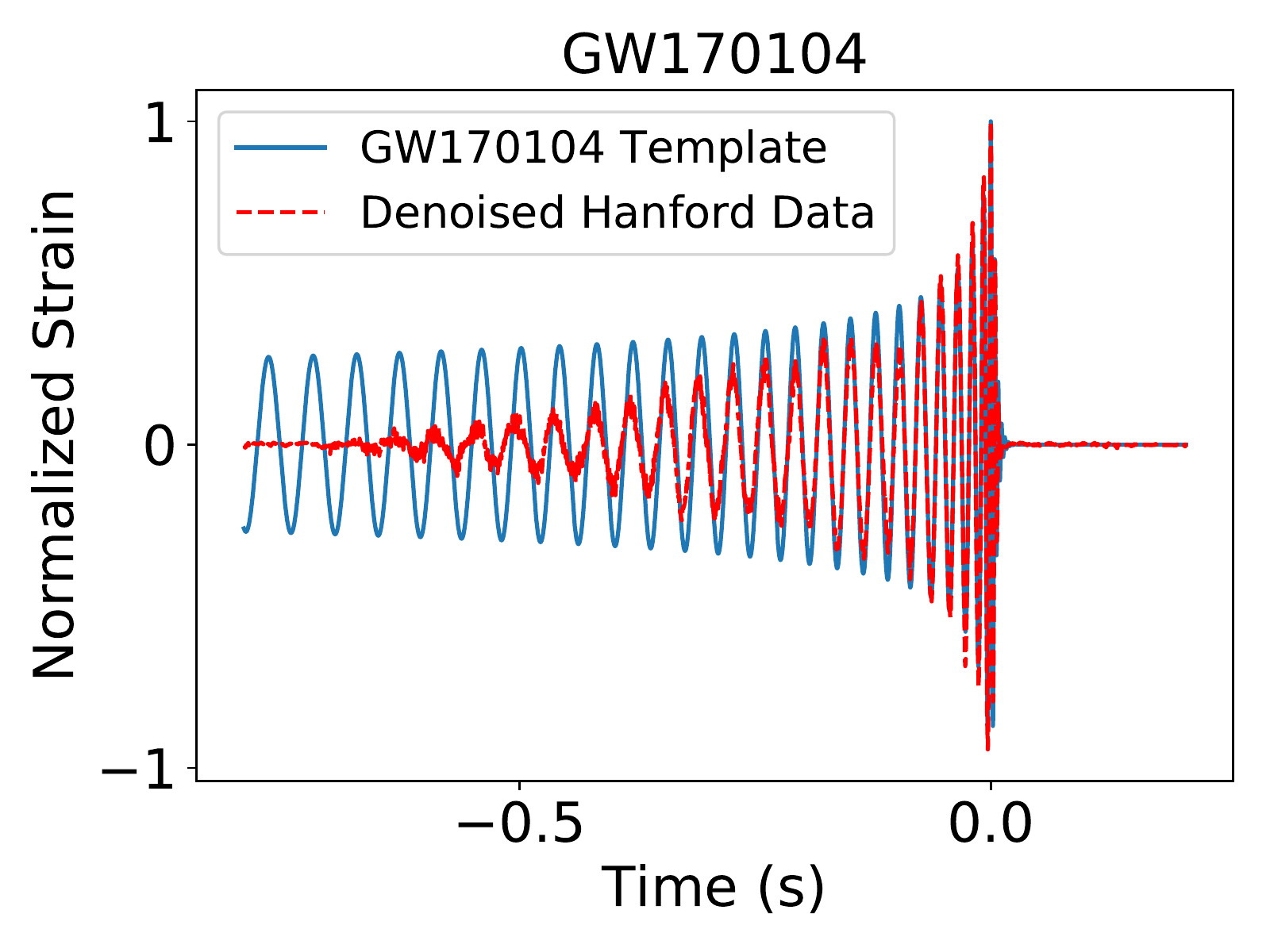}
\includegraphics[width=0.25\linewidth]{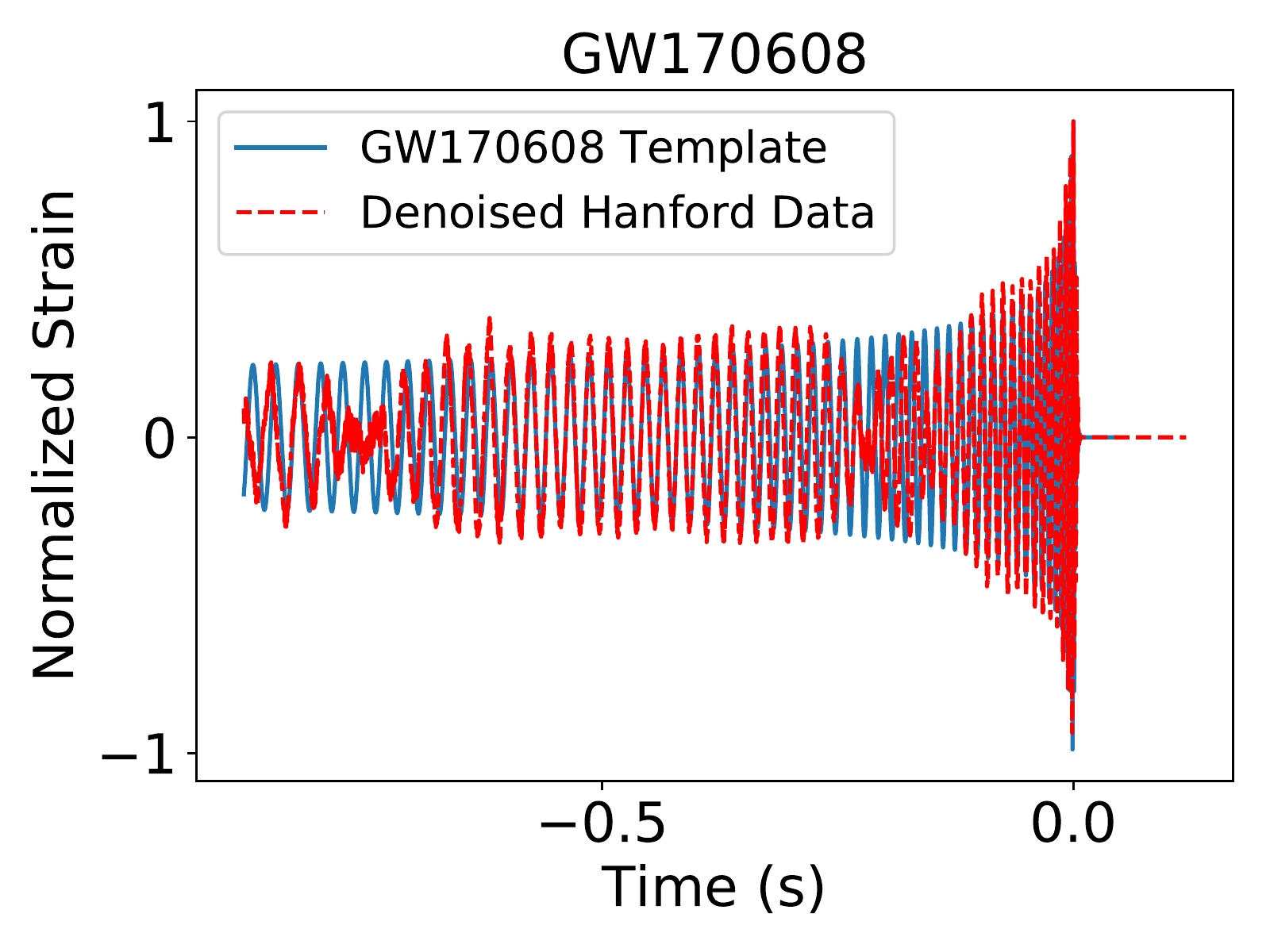}
\includegraphics[width=0.25\linewidth]{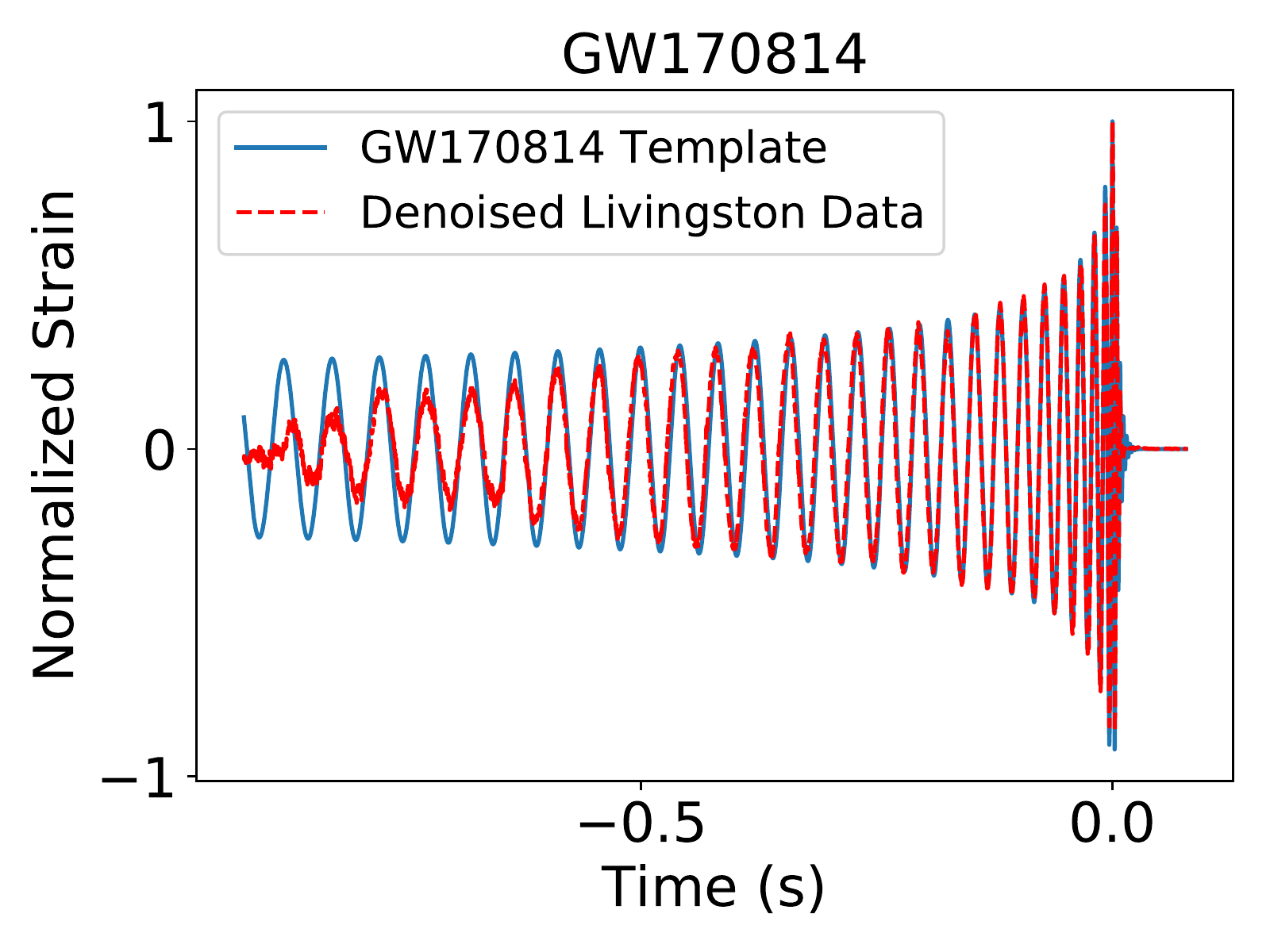}
}
\centerline{
\includegraphics[width=0.25\linewidth]{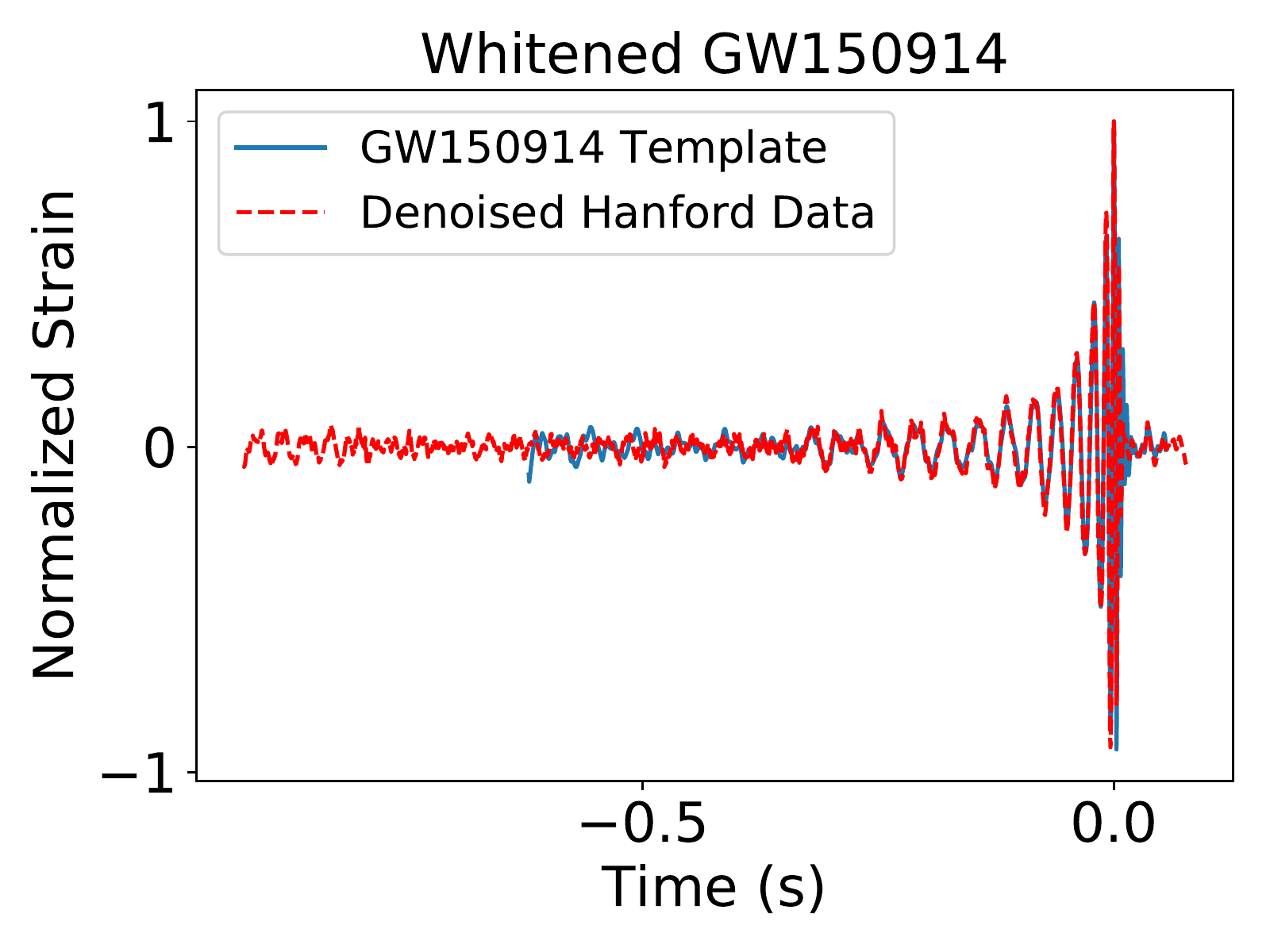}
\includegraphics[width=0.25\linewidth]{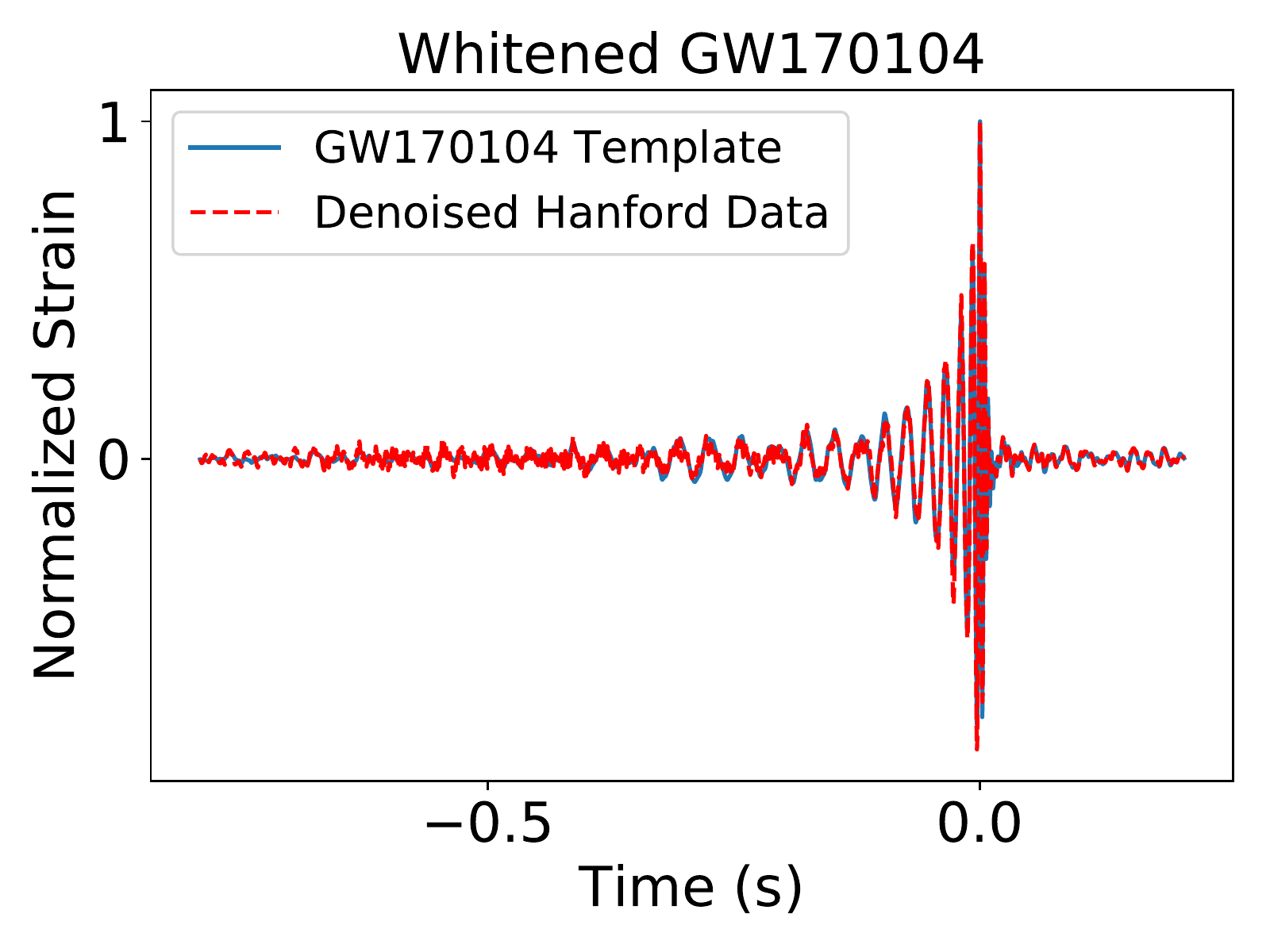}
\includegraphics[width=0.25\linewidth]{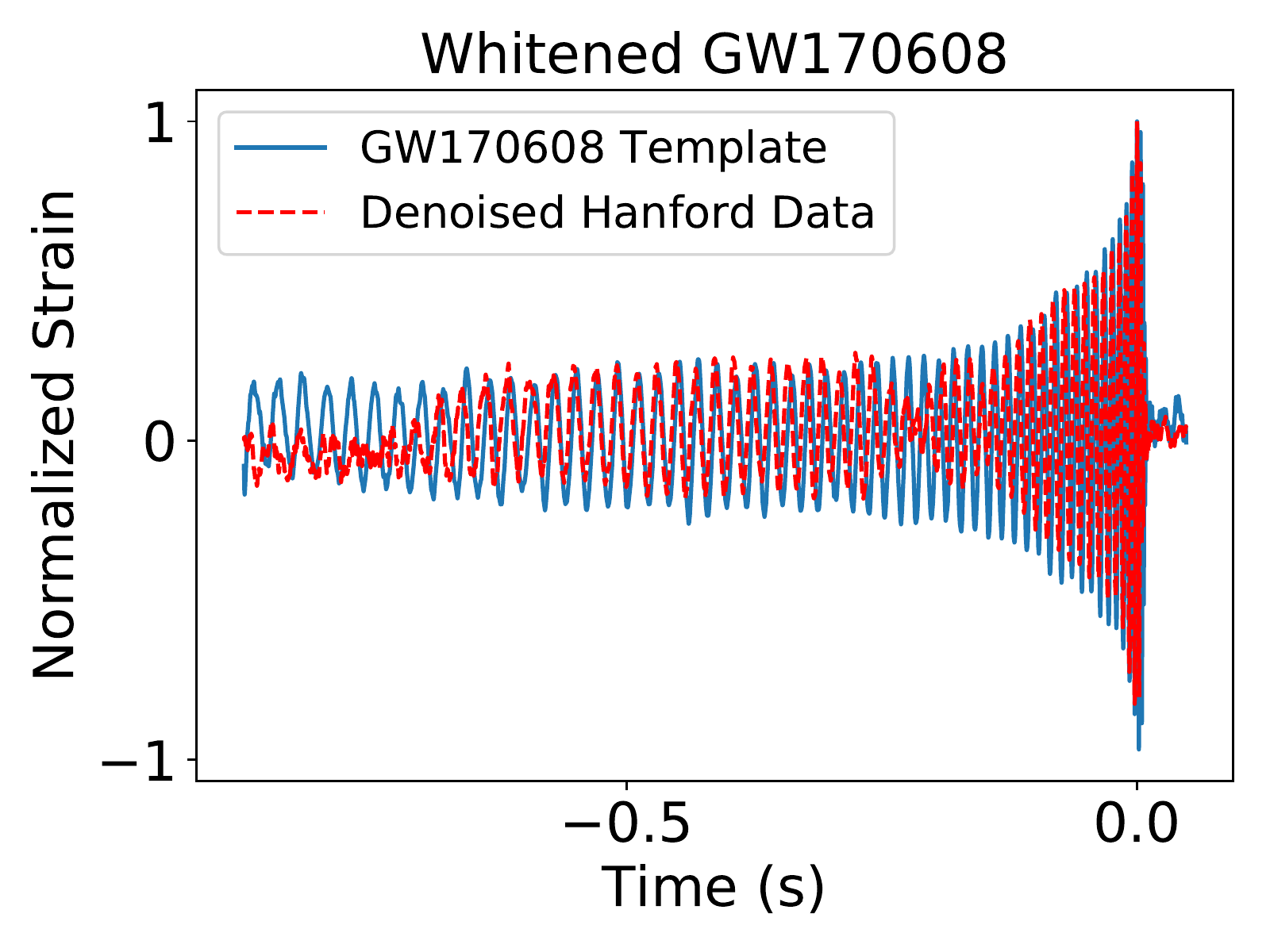}
\includegraphics[width=0.25\linewidth]{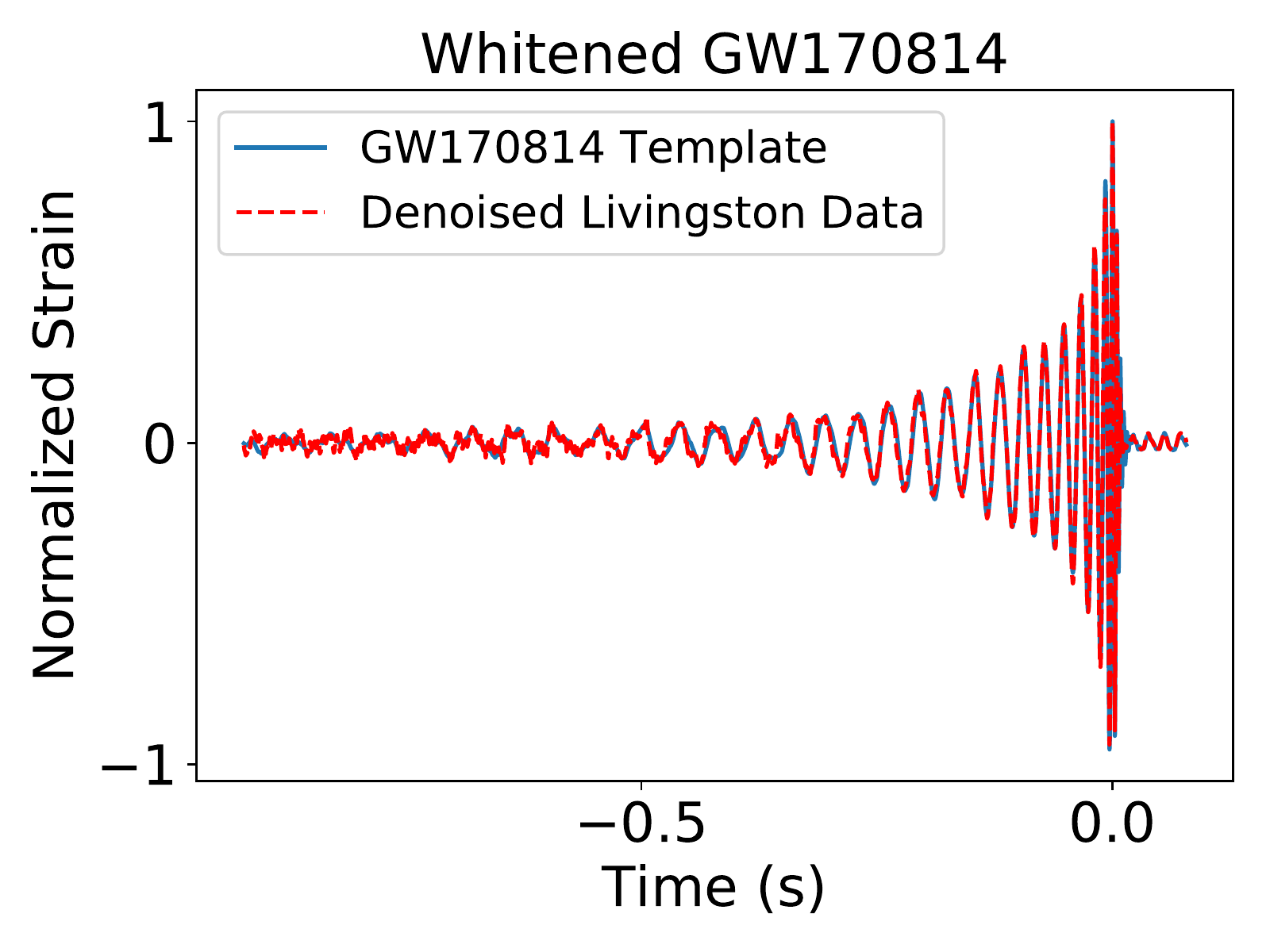}
}
\centerline{
\includegraphics[width=0.25\linewidth]{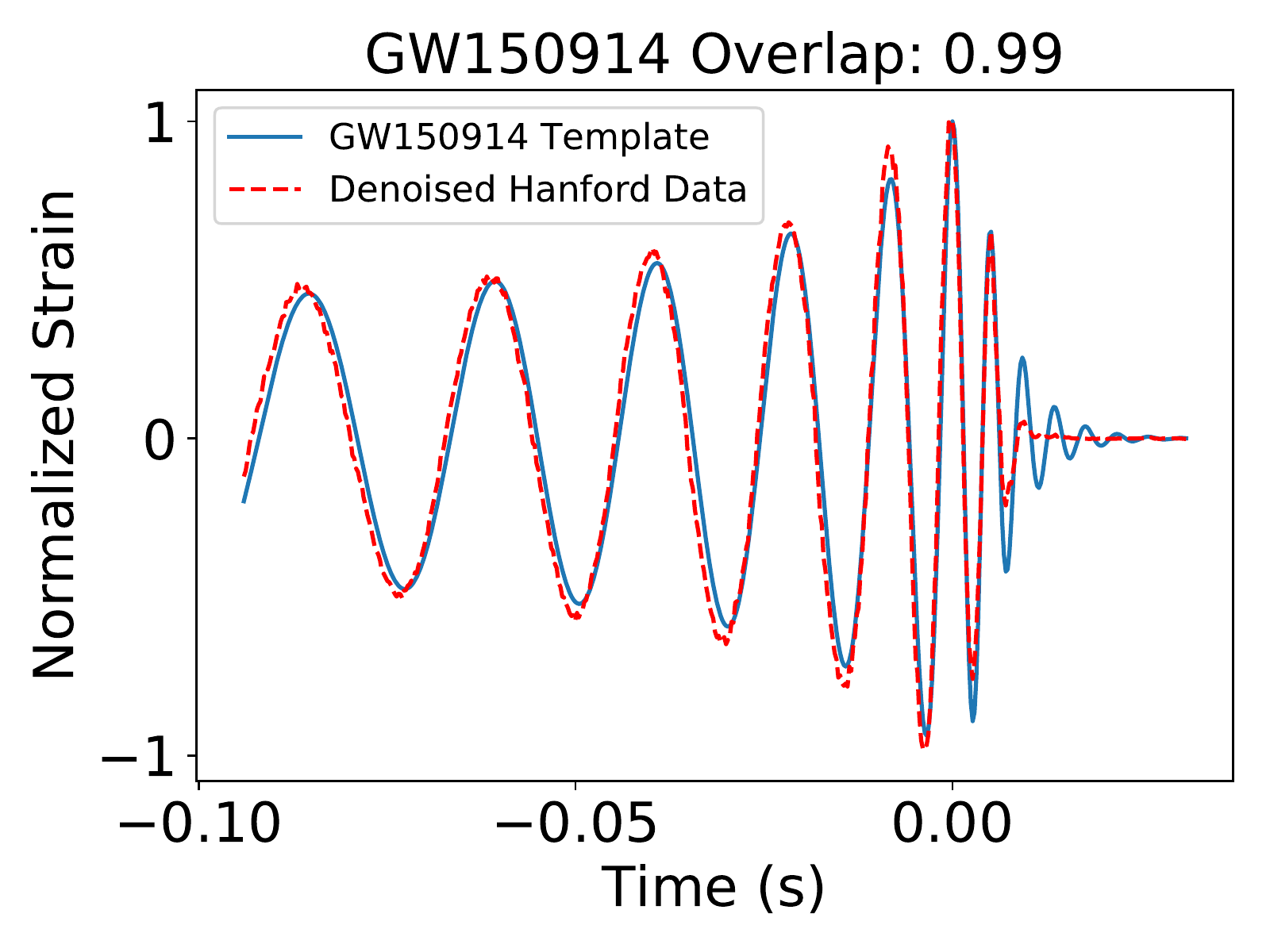}
\includegraphics[width=0.25\linewidth]{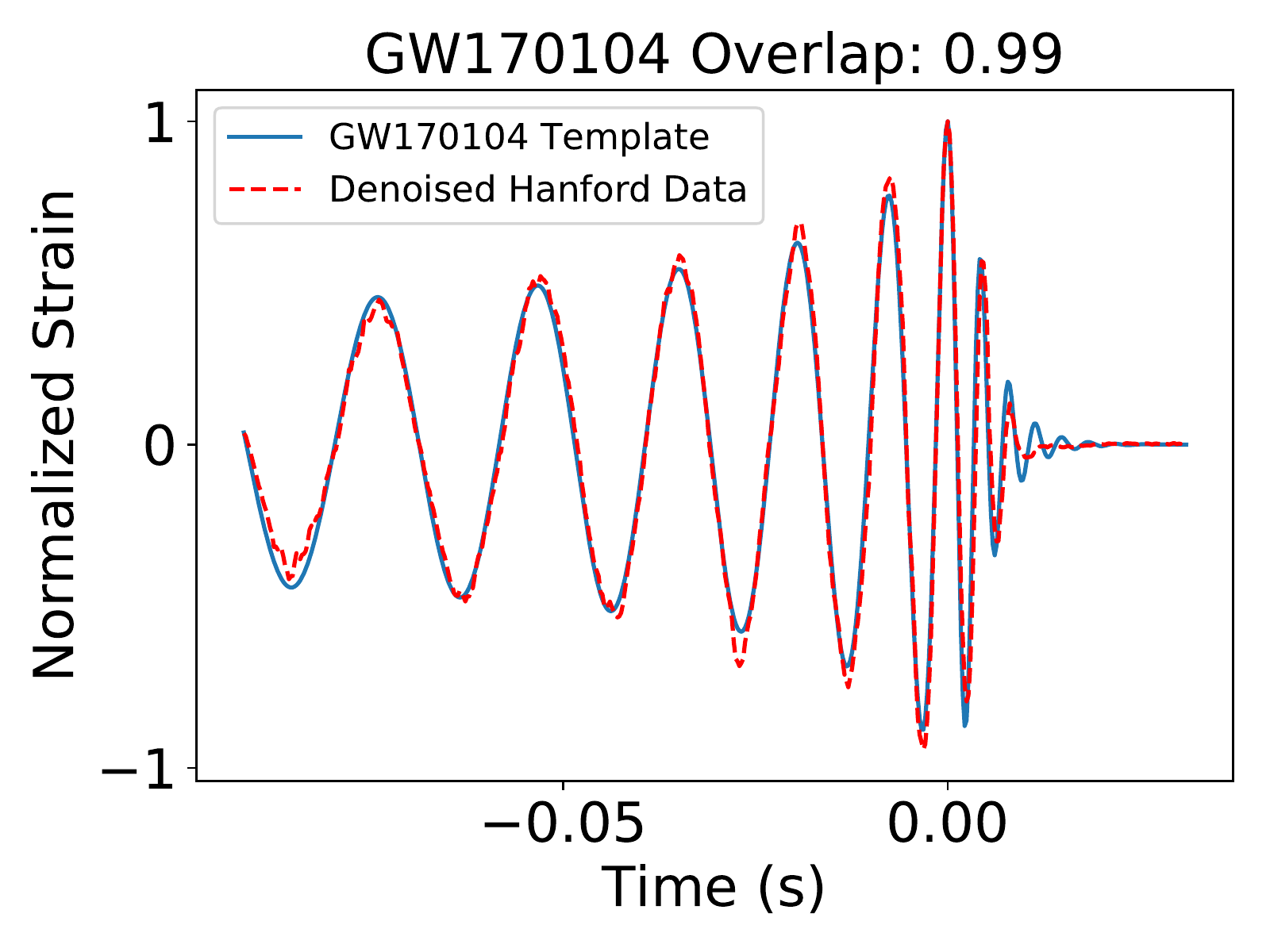}
\includegraphics[width=0.25\linewidth]{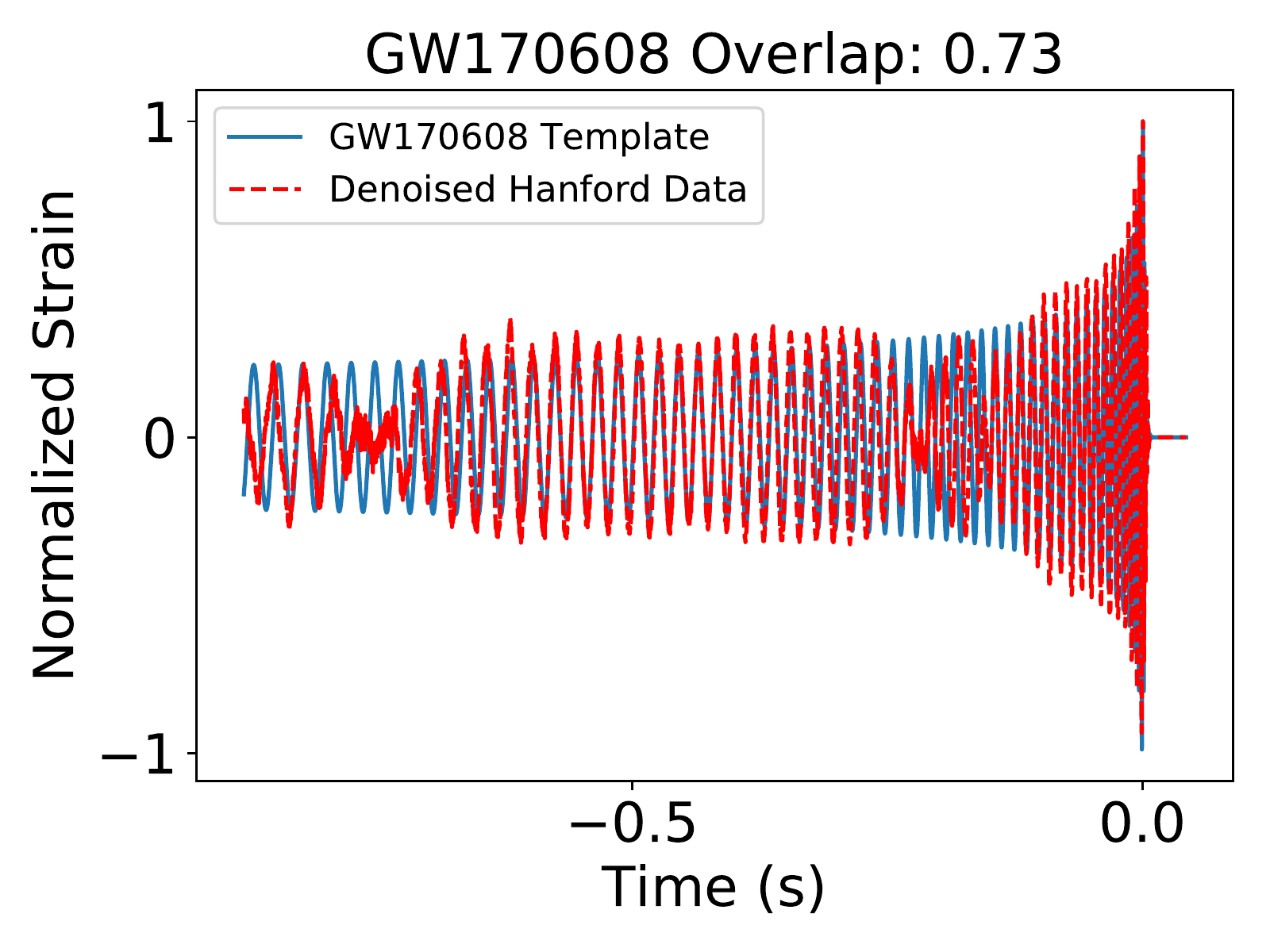}
\includegraphics[width=0.25\linewidth]{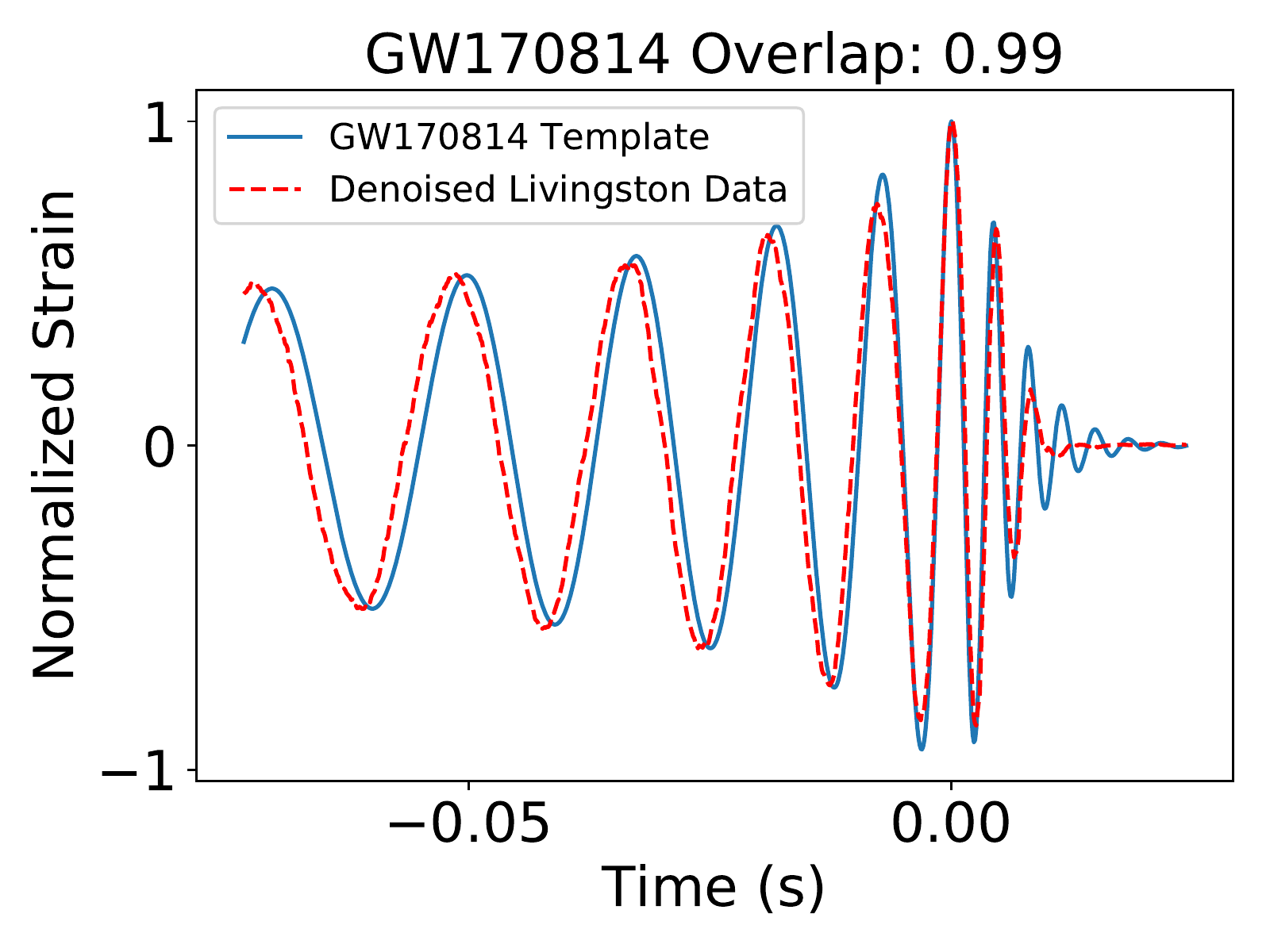}
}
\caption{Top panels (from left to right): denoised signals from the binary black hole mergers GW150914, GW170104, GW170608 and GW170814. Middle panels: overlap between whitened denoised signals and the whitened optimal numerical relativity templates, according to matched-filtering GW detection pipelines~\cite{o1o2catalog,nrsurr:2018K}. Bottom panels: overlap between denoised signals and the optimal numerical relativity templates, according to matched-filtering GW detection pipelines~\cite{o1o2catalog,nrsurr:2018K}.}
\label{real}
\end{figure*}

Figure~\ref{real} presents the output of our denoising algorithm when applied to real advanced LIGO noise that contains four different BBH events. We distill these set of results in two cases. The first one comprises GW signals that describe the GW events GW150914, GW170104 and GW170814. To denoise these GW signals, we fed into our denoising algorithm a one second-long advanced LIGO data segment that contains the GW under consideration. The output of our denoising algorithm for each of these events is shown in the top panels of Figure~\ref{real}. It is important to clearly delineate the realm of applicability of our results, since our deep learning algorithm provides a realistic description of the data when the GW signal is actually detectable by advanced LIGO. Thus, to clearly exhibit the portion of the denoised data that may be used for data analysis studies the mid-panels of Figure~\ref{real} show the whitened true GW signal and the whitened output of our denoising algorithm. The bottom panels in this Figure show the overlap between the output of our denoising algorithm, within its realm of applicability, and the NR templates that optimally describe these signals~\cite{o1o2catalog,nrsurr:2018K}. We notice that in all cases \({\cal{O}}\geq0.99\). We have selected these systems to consider a broad range of masses, mass-ratios and SNRs of recently detected BBH mergers. These results show that deep learning can provide unwhitened time-series data which may facilitate rapid analyses to constrain the parameter space that describes BBH mergers.

\begin{figure}
\centering
\includegraphics[width=0.8\linewidth]{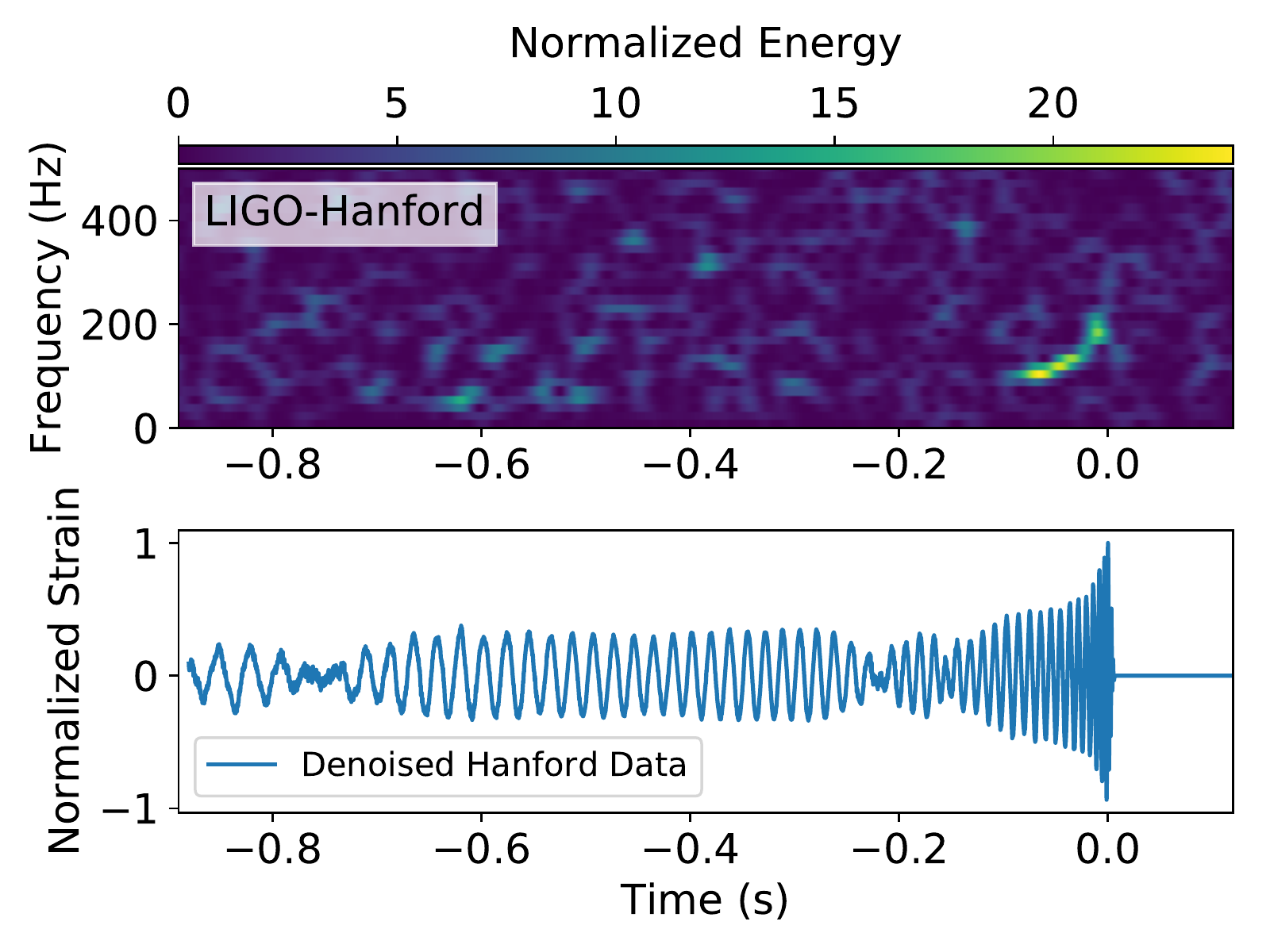}
\caption{Top panel: time-frequency representation of the GW170608 signal. Bottom panel: time-series denoised signal produced by our deep learning algorithm. Notice that the reconstruction of the denoised signal is determined by the loudness of the signal and the sensitivity of the detector.}
\label{spec}
\end{figure}

\noindent There are several important aspects of our denoising algorithm that we want to exhibit with the GW event GW170608. This is a low mass BBH merger with moderate SNR~\cite{GW170608}. As shown in the time-frequency power maps of the LIGO strain data produced for this event in Figure~(1) at~\cite{GW170608}, and the spectrogram we have produced with data available at the Gravitational Wave Open Science Center~\cite{losc} in Figure~\ref{spec}, we notice that the characteristic chirping morphology of the BBH evolution is rather intermittent, as opposed to the smooth, continuous time-frequency tracks observed in other GW events~\cite{DI:2016}. Therefore, based on the loudness and low total mass of GW170608, we would expect to observe these signatures in the output time-series data of our denoising algorithm. Our results, presented in the bottom panel of Figure~\ref{spec}, and the third bottom panel (from left to right) in Figure~\ref{real}, demonstrate that this is exactly what we observe in our denoised GW170608 signal, namely, at lower frequencies our denoised signal stays in phase with the optimal NR template that, according to matched-filtering algorithms, reproduces GW170608~\cite{o1o2catalog,nrsurr:2018K}. As the BBH system nears merger, the power of the signal drops significantly, which is reflected in the reconstruction of our denoised signal. During this time window, localized around \(t\sim-0.2~\textrm{seconds}\) in the bottom panel of Figure~\ref{spec}, our denoised signal goes out of phase and amplitude with the NR template. Right before merger, the true GW signal increases its SNR and our denoised signal is now reconstructed with high fidelity.

The above description is essential to highlight that our denoising algorithm has not just hierarchically learned the properties of GW signals, and then performed an interpolation of these abstract features to produce a denoised signal. Rather, our denoising algorithm is actually using the statistics of the noise in which the signal is embedded to provide a realistic representation of the true GW event. This reconstruction is determined by the SNR of the signal, and encodes the sensitivity of the detectors at the time of observation. 

\noindent These results furnish additional evidence for the versatility and power of deep learning for GW data analysis in realistic detection scenarios, and represent the first time deep learning is proven effective at denoising BBH GW signals that span a broad range of SNRs. 

\subsection{Spin-precessing binary black hole waveforms}

As described in Section~\ref{sec:curation}, we trained our deep learning algorithm to remove noise from waveform signals that describe non-spinning BBH mergers. In the previous section we have demonstrated that our denoising algorithm generalizes to new types of signals, since some of the BBH waveforms that we denoised are consistent with BBHs that have non-zero spins, as shown in~\cite{CatalogBBH:2018}. In this section, we quantify the robustness of our denoising algorithm in a more challenging scenario, i.e., we consider spin-precessing BBH mergers, produced with the waveform model introduced in~\cite{seobnrv3}, with the following parameters: total mass \(M=\{70\msun,\,75\msun\}\), mass-ratio \(q=\{4/3,\,4\}\), and three spin-vector combinations, namely:

\begin{itemize}
    \item \(\mathbf{\hat{s}_1}=\,(0.2,\,0.3,\,0.5)\), \(\mathbf{\hat{s}_2}=(0.3,\,-0.4,\,0.5)\),
    \item \(\mathbf{\hat{s}_1}=(0.4,\,0.5,\,0.7)\), \(\mathbf{\hat{s}_2}=(0.4,\,-0.5,\,0.7)\),
    \item \(\mathbf{\hat{s}_1}=\left(0.5,\,0.5,\,0.7\right)\), \(\mathbf{\hat{s}_2}=(0.5,\,0.5,\,-0.7)\).
\end{itemize}

\noindent These spin-precessing BBH waveforms are whitened using a noise PSD estimate in the vicinity of the event GW150914. We then extract open source advanced LIGO data around this same event and inject them therein (note that in all these studies the data containing the true BBH GW signals are excised).

\begin{figure*}
\centerline{
\includegraphics[width=0.33\linewidth]{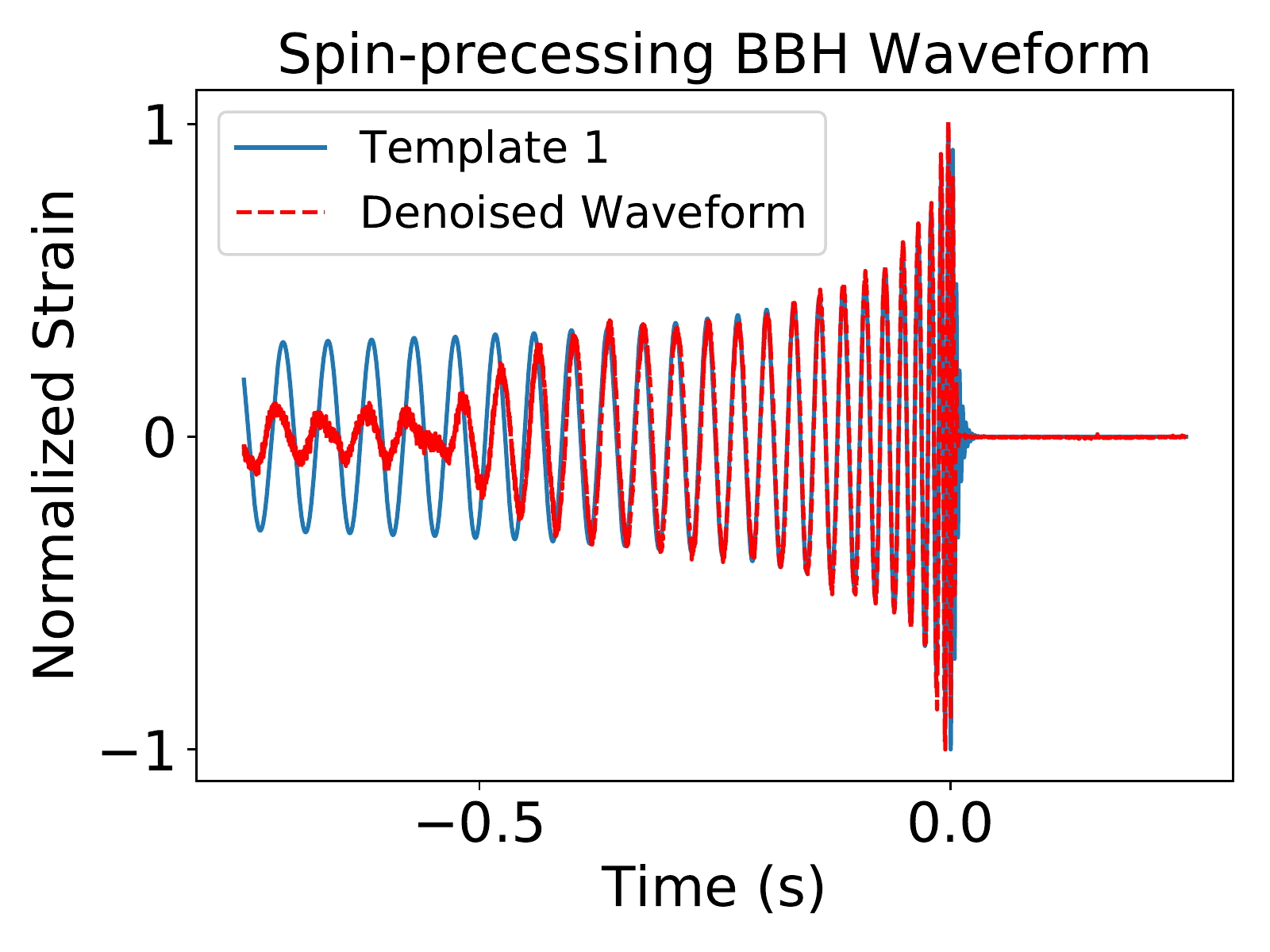}
\includegraphics[width=0.33\linewidth]{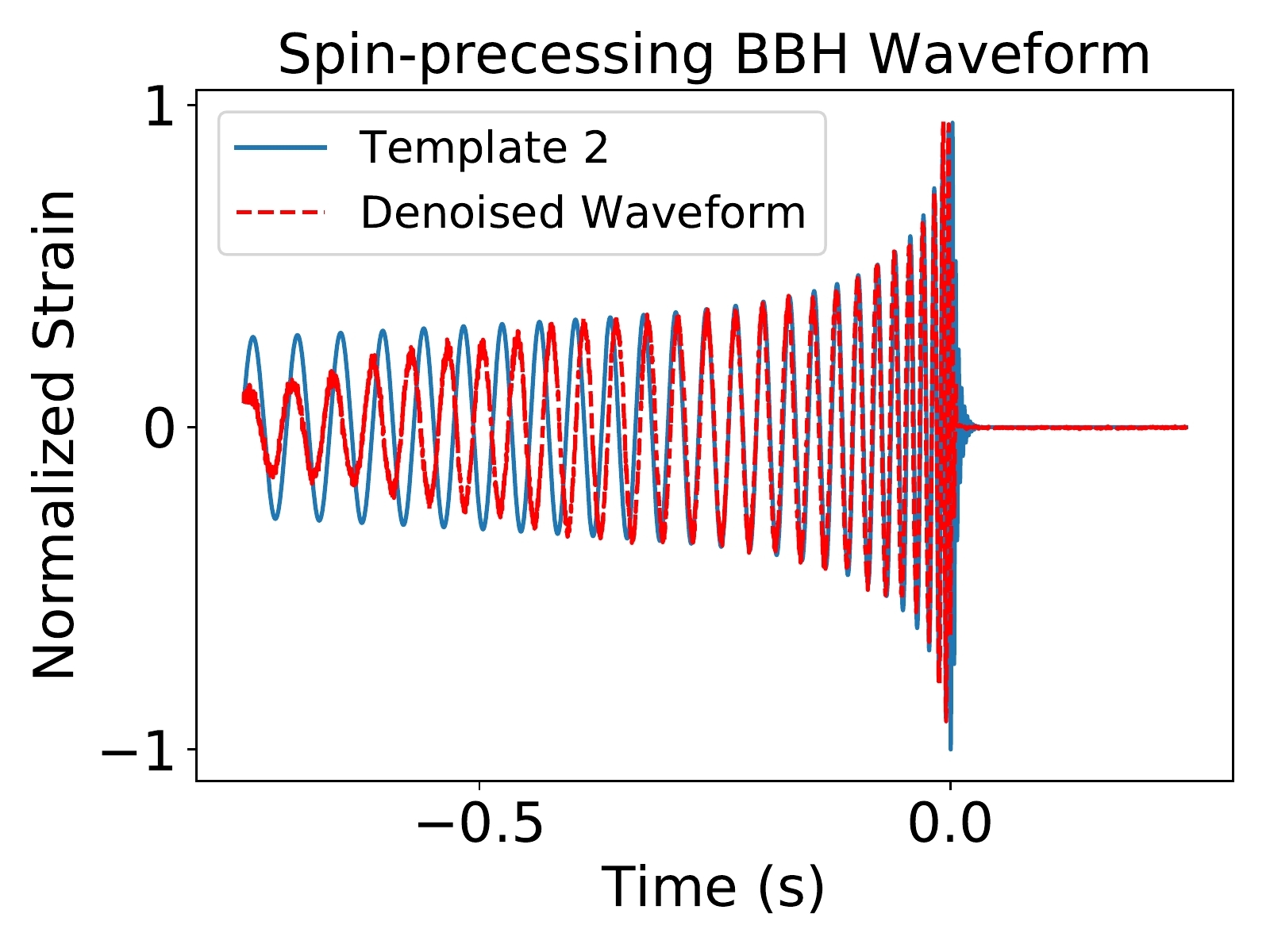}
\includegraphics[width=0.33\linewidth]{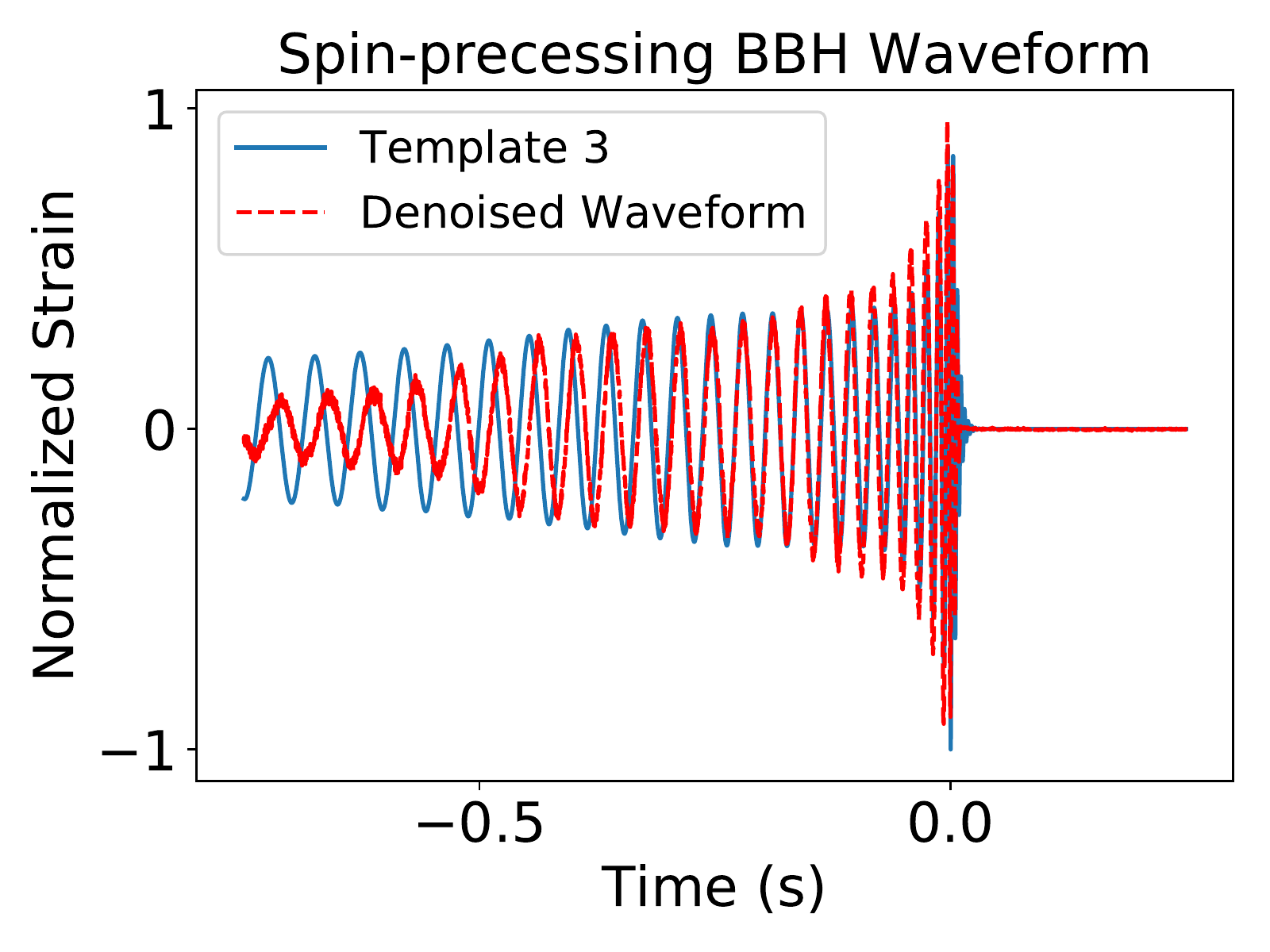}
}
\centerline{
\includegraphics[width=0.33\linewidth]{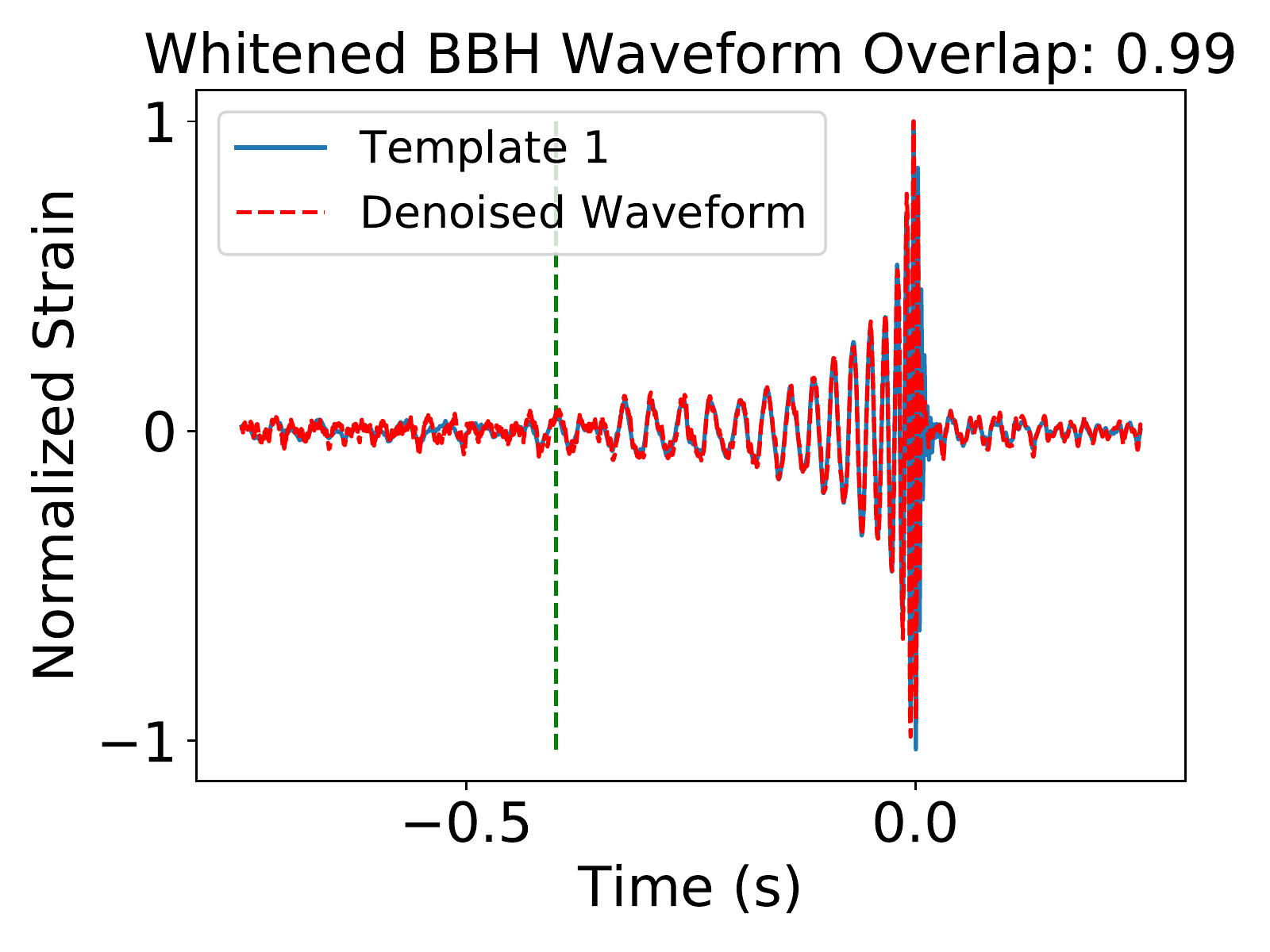}
\includegraphics[width=0.33\linewidth]{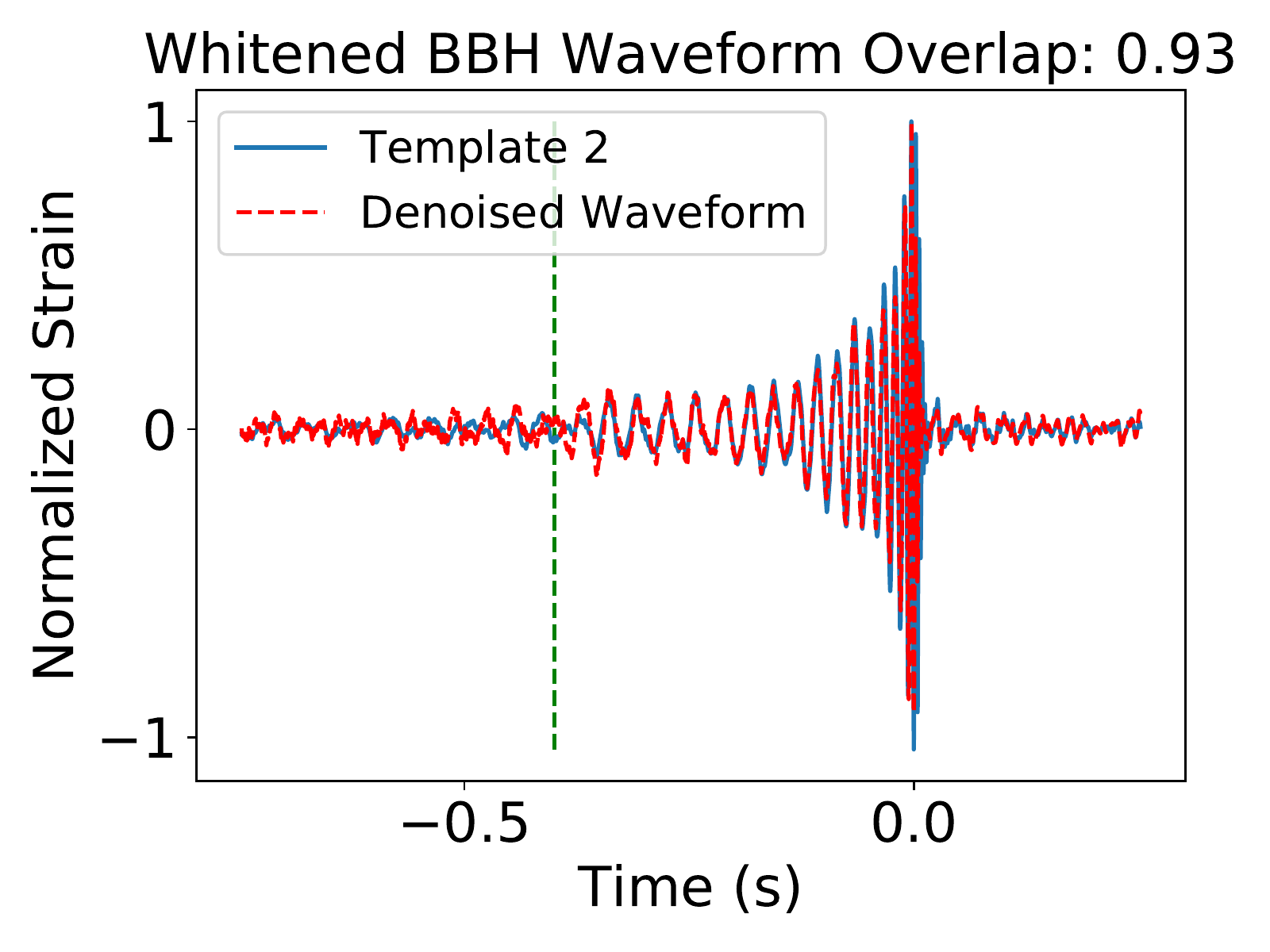}
\includegraphics[width=0.33\linewidth]{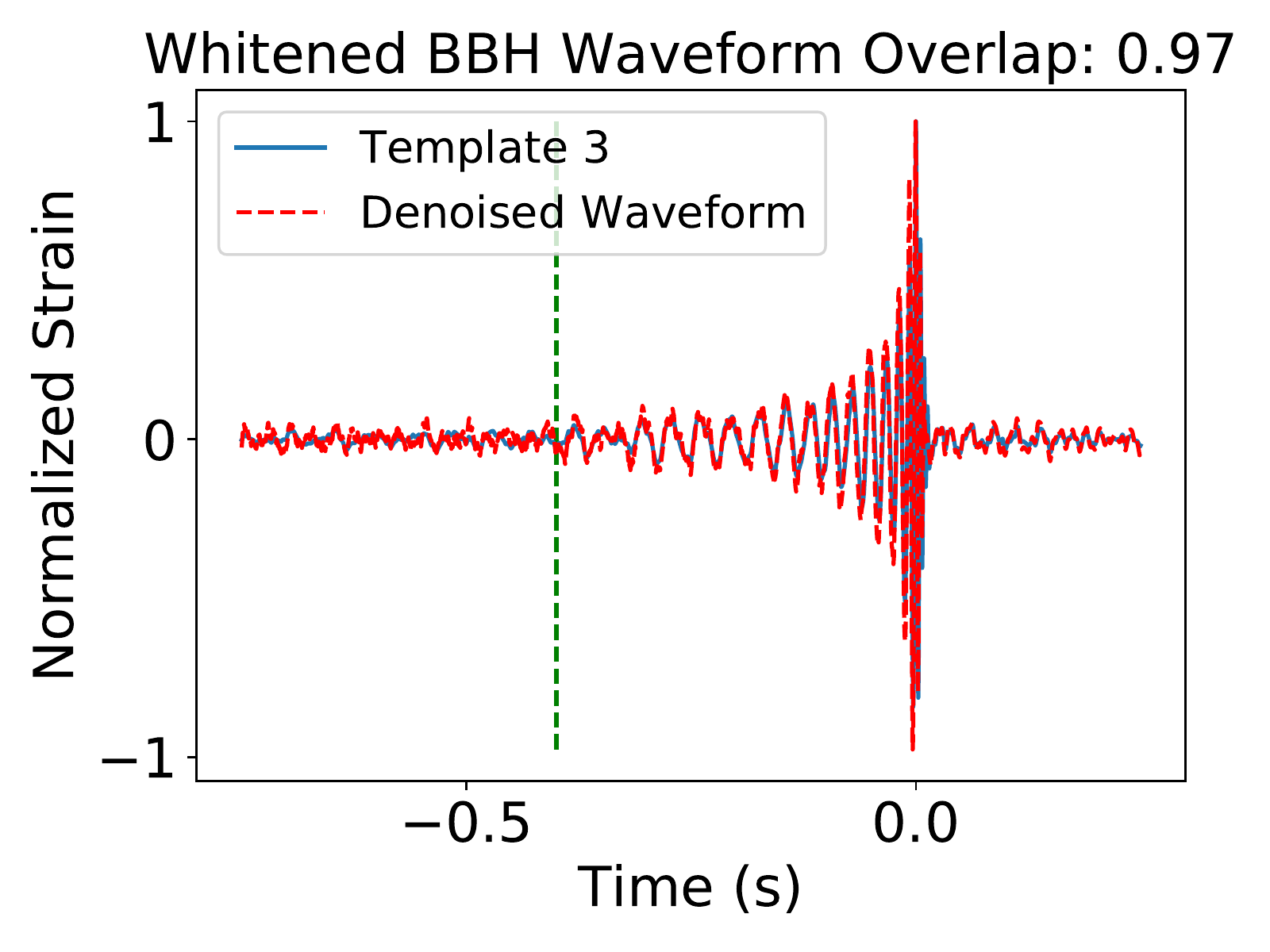}
}
\caption{Denoising of three gravitational wave signals, embedded in real advanced LIGO noise, that describe spin-precessing binary black hole mergers with component masses \((m_1,\,m_2)\)= \((40\msun,\,30\msun)\). Left panel: the 3-D spin vectors of the binary components are \(\mathbf{\hat{s}_1}=\,(0.2,\,0.3,\,0.5)\), \(\mathbf{\hat{s}_2}=(0.3,\,-0.4,\,0.5)\); mid panel: 
    \(\mathbf{\hat{s}_1}=(0.4,\,0.5,\,0.7)\), \(\mathbf{\hat{s}_2}=(0.4,\,-0.5,\,0.7)\); right panel: 
    \(\mathbf{\hat{s}_1}=\left(0.5,\,0.5,\,0.7\right)\), \(\mathbf{\hat{s}_2}=(0.5,\,0.5,\,-0.7)\). All templates have matched-filtering \(
\textrm{SNR}=13\).}
\label{fig:spin_pres}
\end{figure*}

Figure~\ref{fig:spin_pres} presents two sets of results: (i) the top panels show the unwhitened time-series data of the ground-truth waveform and the output of our denoising algorithm; (ii) as discussed before, it is important to explicitly show the realm of applicability of the output time-series data of our deep learning algorithm, and to inform that we show in the bottom panels the whitened versions of the ground-truth signals and the denoised signals. These panels also show the overlap between the ground-truth signals and the denoised waveforms, computed from the time marked by the dashed lines to the last time-sample of the signals. These panels show, from left to right, low-, moderate- and high-spin configurations. The overlap values reported in these panels, \({\cal{O}}=\{0.99,\,0.93,\,0.97\}\) indicate that, even though we trained our deep learning denoiser with non-spinning BBH signals, we can still reconstruct the features of spin-precessing BBH mergers. We have also quantified the ability of our denoising algorithm to generalize to these new types of signals by computing the overlap between these spin-precessing signals and the entire dataset of waveforms we used to train our denoising algorithm. The corresponding overlap values for the signals shown in Figure~\ref{fig:spin_pres}, from left to right, are \({\cal{O}}=\{0.93,\, 0.83,\, 0.94\}\). If we compare these results with the actual overlap values obtained between the denoised signals and the ground-truth spin-precessing waveforms, i.e., \({\cal{O}}=\{0.99,\,0.93,\,0.97\}\),  we realize that the denoiser has been able to generalize to new types of signals that are not present in the training dataset.

\begin{figure*}
\centerline{
\includegraphics[width=0.33\linewidth]{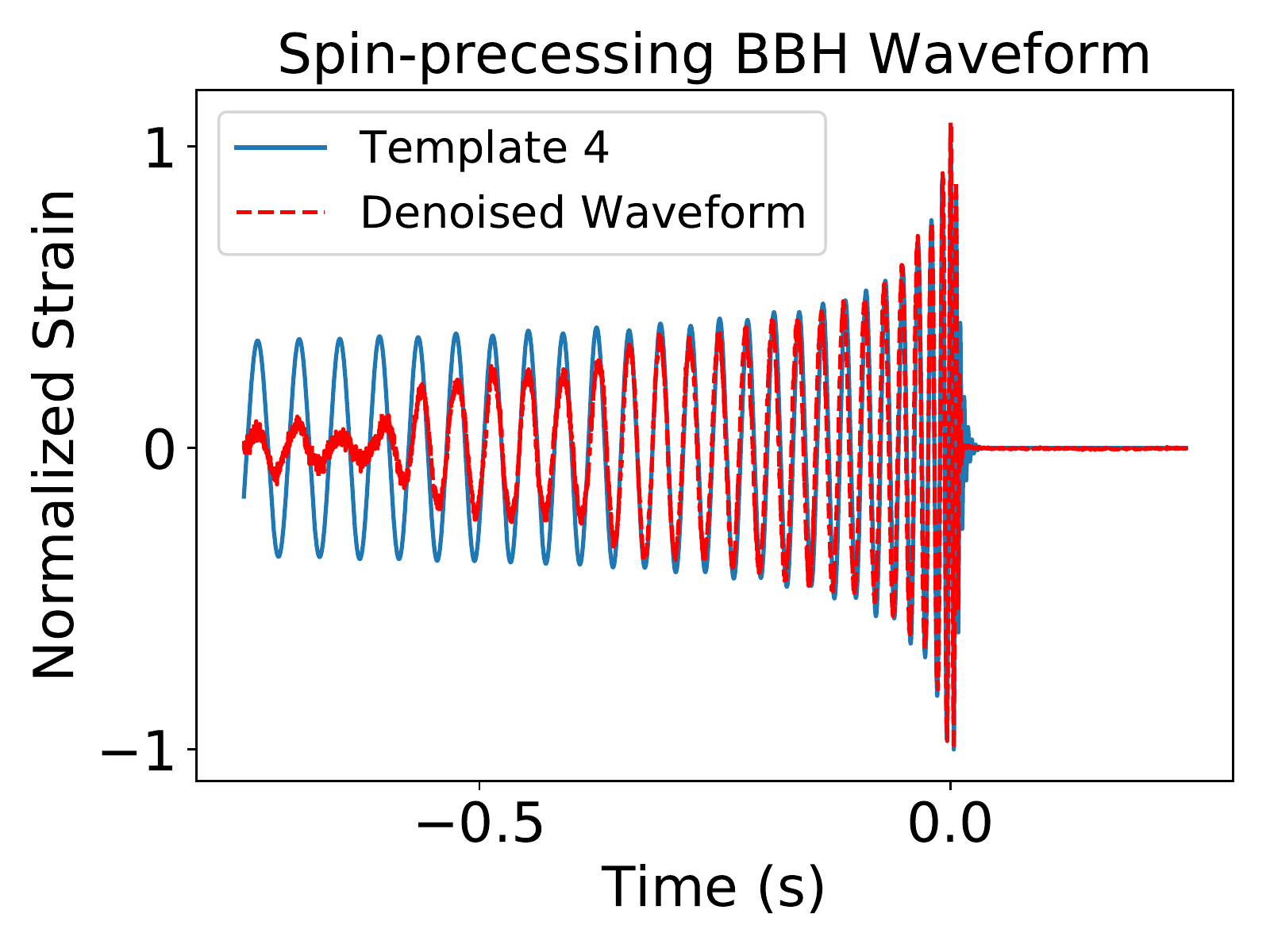}
\includegraphics[width=0.33\linewidth]{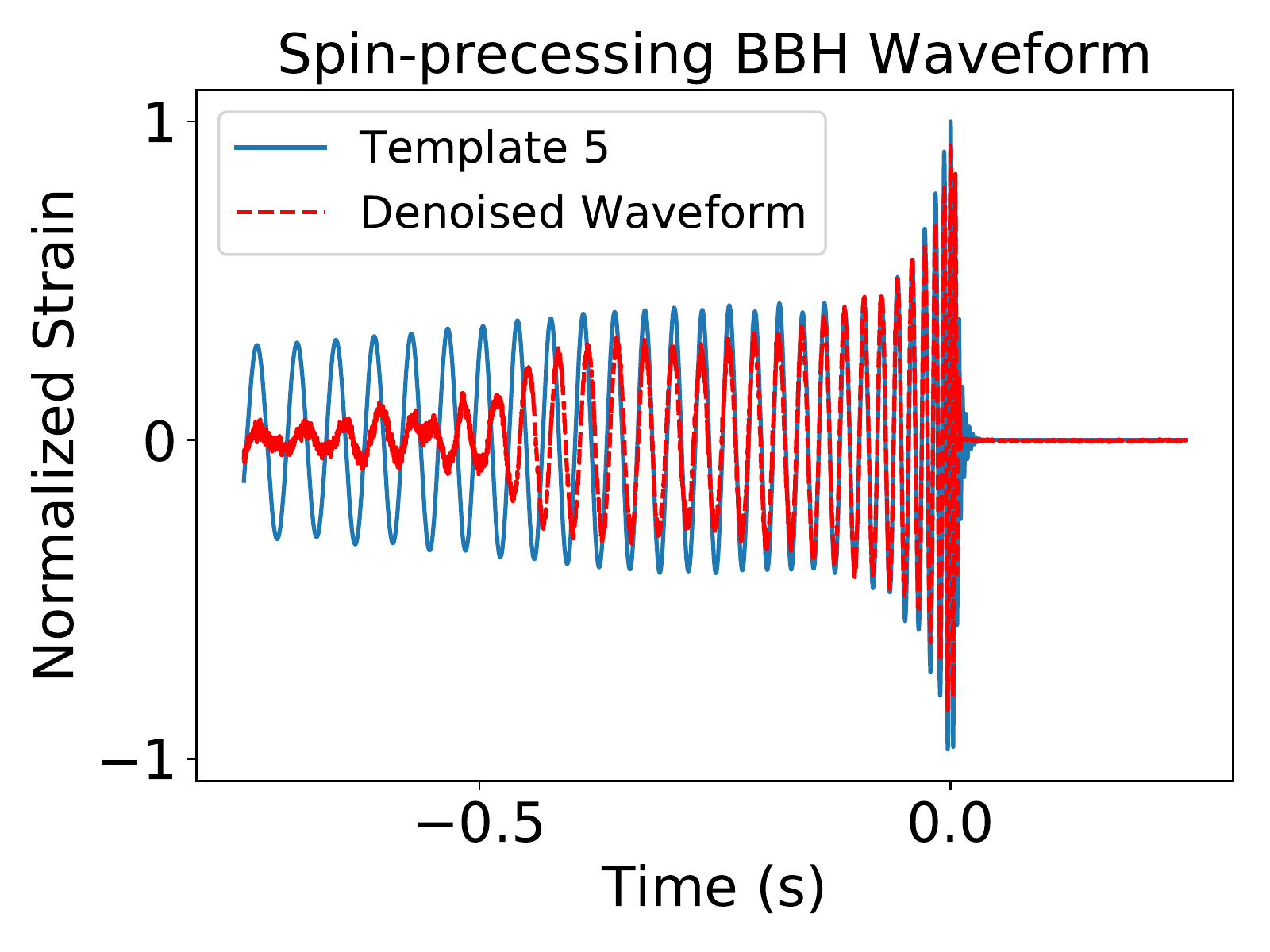}
\includegraphics[width=0.33\linewidth]{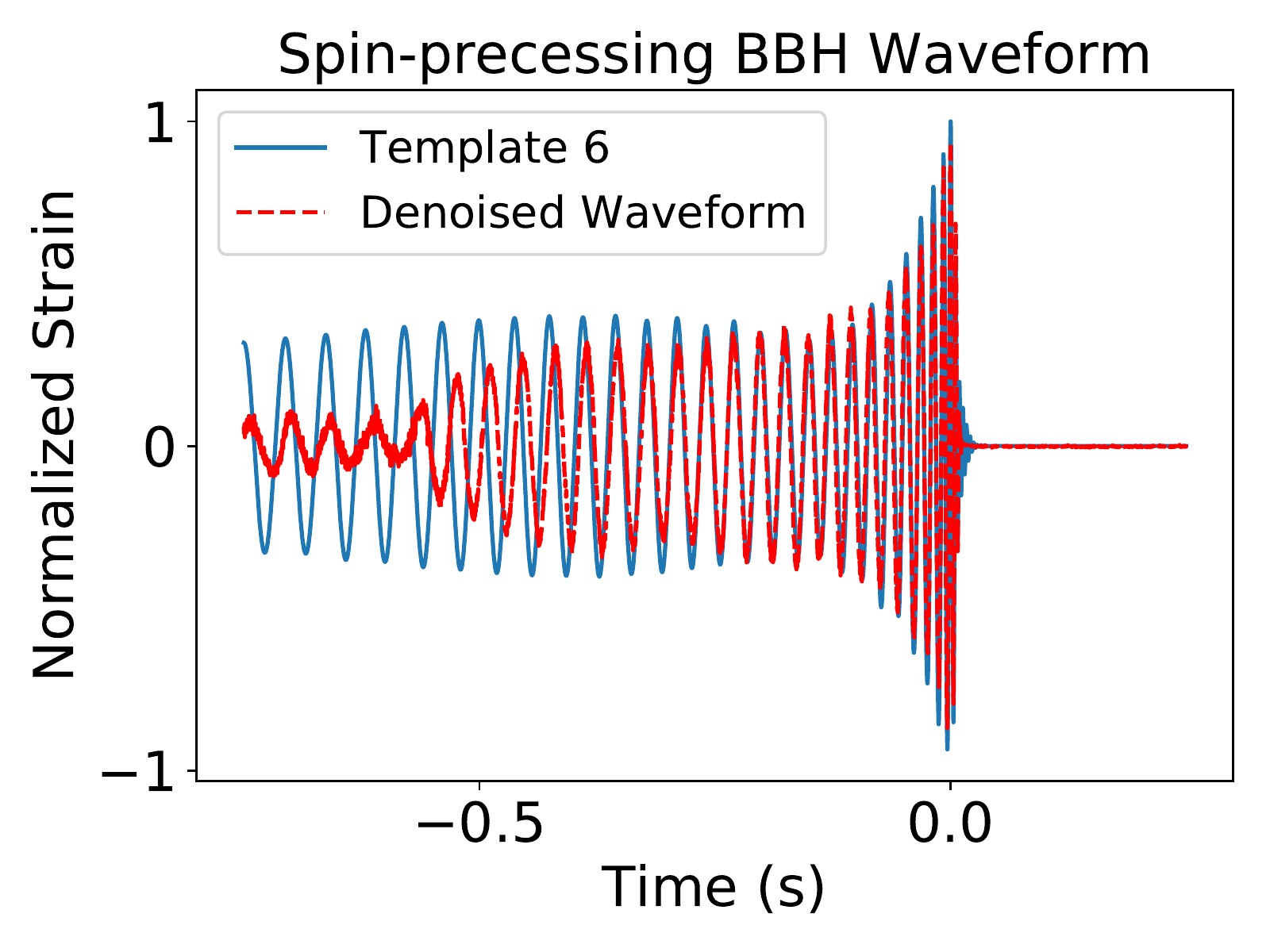}
}
\centerline{
\includegraphics[width=0.33\linewidth]{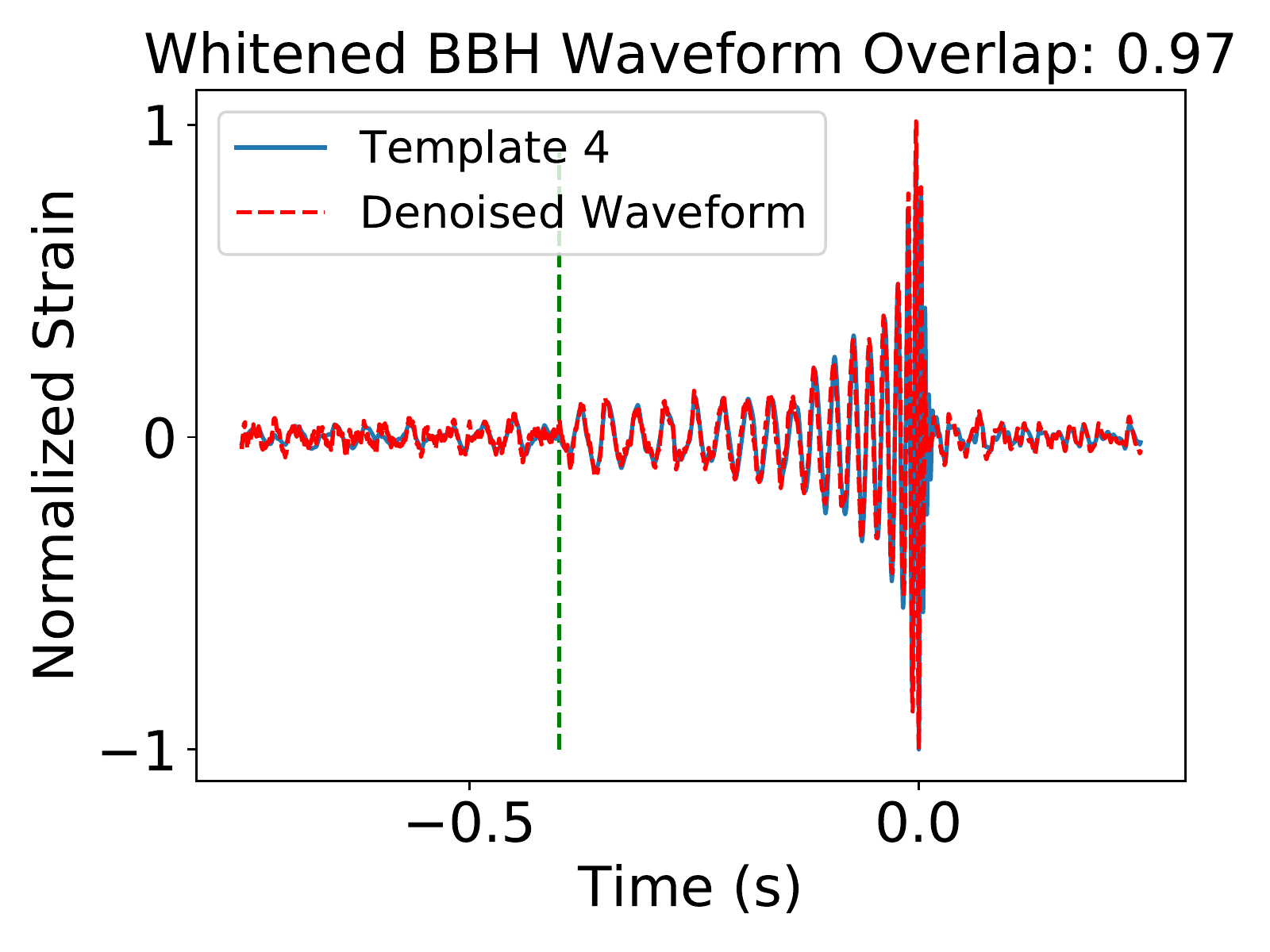}
\includegraphics[width=0.33\linewidth]{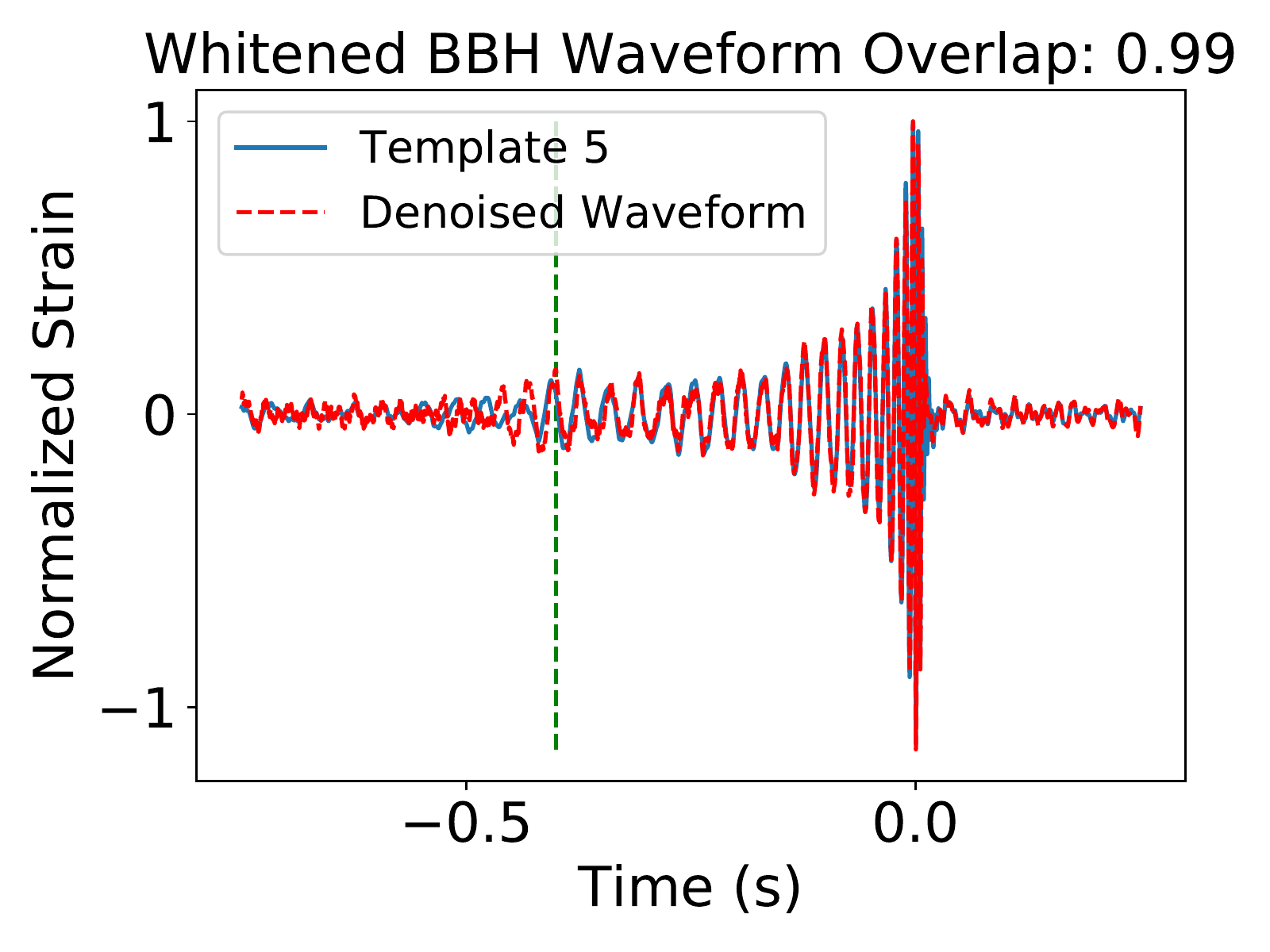}
\includegraphics[width=0.33\linewidth]{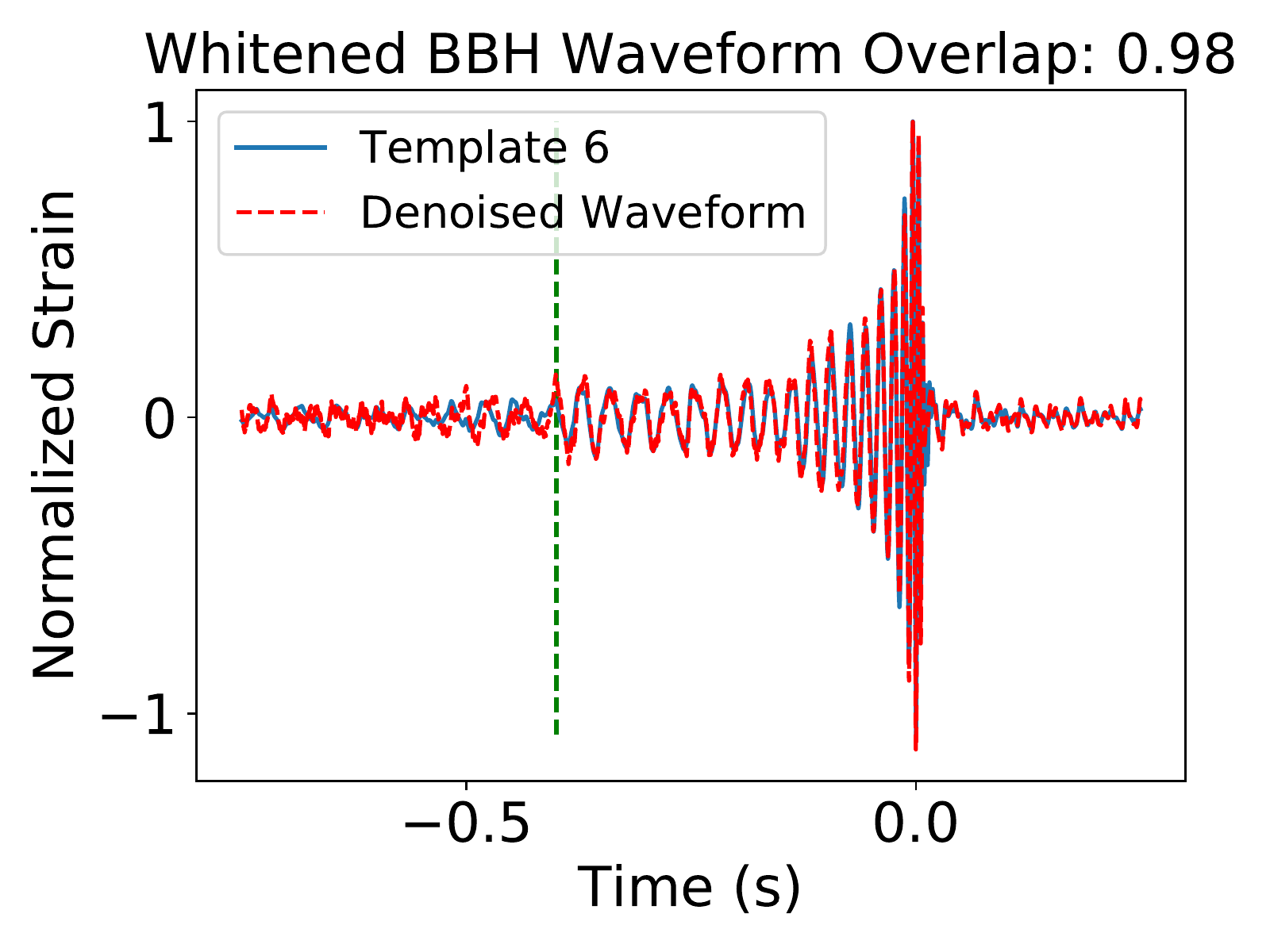}
}
\caption{As Figure~\ref{fig:spin_pres}, but now for binary black hole signals with total mass \(M=75\msun\) and mass-ratio \(q=4\). The spin vectors are the same as those reported in Figure~\ref{fig:spin_pres}.}
\label{fig:spin_pres_two}
\end{figure*}

In Figure~\ref{fig:spin_pres_two} we perform another analysis concerning the suitability of our deep learning algorithm to denoise spin-precessing BBH signals that exhibit more clearly the features of spin-precession. To do so we have chosen a systems with mass-ratio \(q=4\). The three panels in Figure~\ref{fig:spin_pres_two} present, from left to right, the overlaps between the denoised time-series signals and the ground-truth signals, namely \({\cal{O}}=\{0.97,\,0.99,\,0.98\}\). As before, we have also computed the overlap between these spin-precessing signals and the entire data set of waveforms used to train our deep learning algorithm, finding that the corresponding overlaps for the signals shown in Figure~\ref{fig:spin_pres_two}, from left to right, are \({\cal{O}}=\{0.91,\, 0.83,\, 0.93\}\).

These analyses furnish evidence that our denoising algorithm can generalize to new types of signals, and also sheds light on regions of parameter space where our algorithms requires additional work, in particular spin-precessing BBHs with asymmetric mass-ratios. Informed by these findings, we will present an extended version of this algorithm in future work to recover with higher fidelity these type of astrophysical events. For now, it is worth highlighting that this method can be readily applied to LIGO data analysis given the measured spin values of detected BBH mergers~\cite{CatalogBBH:2018}.

\subsection{Glitches}
\label{sec:glitches}

An important consideration in the construction of denoising algorithms is that they should be trained so as to tell apart noise anomalies from true signals. Therefore, a metric to assess whether the denoising algorithm performs optimally consists of contaminating a given BBH waveforms with a variety of glitches and ensure that the denoised signals are not altered by them.

\begin{figure*}
\centerline{
\includegraphics[width=0.5\linewidth]{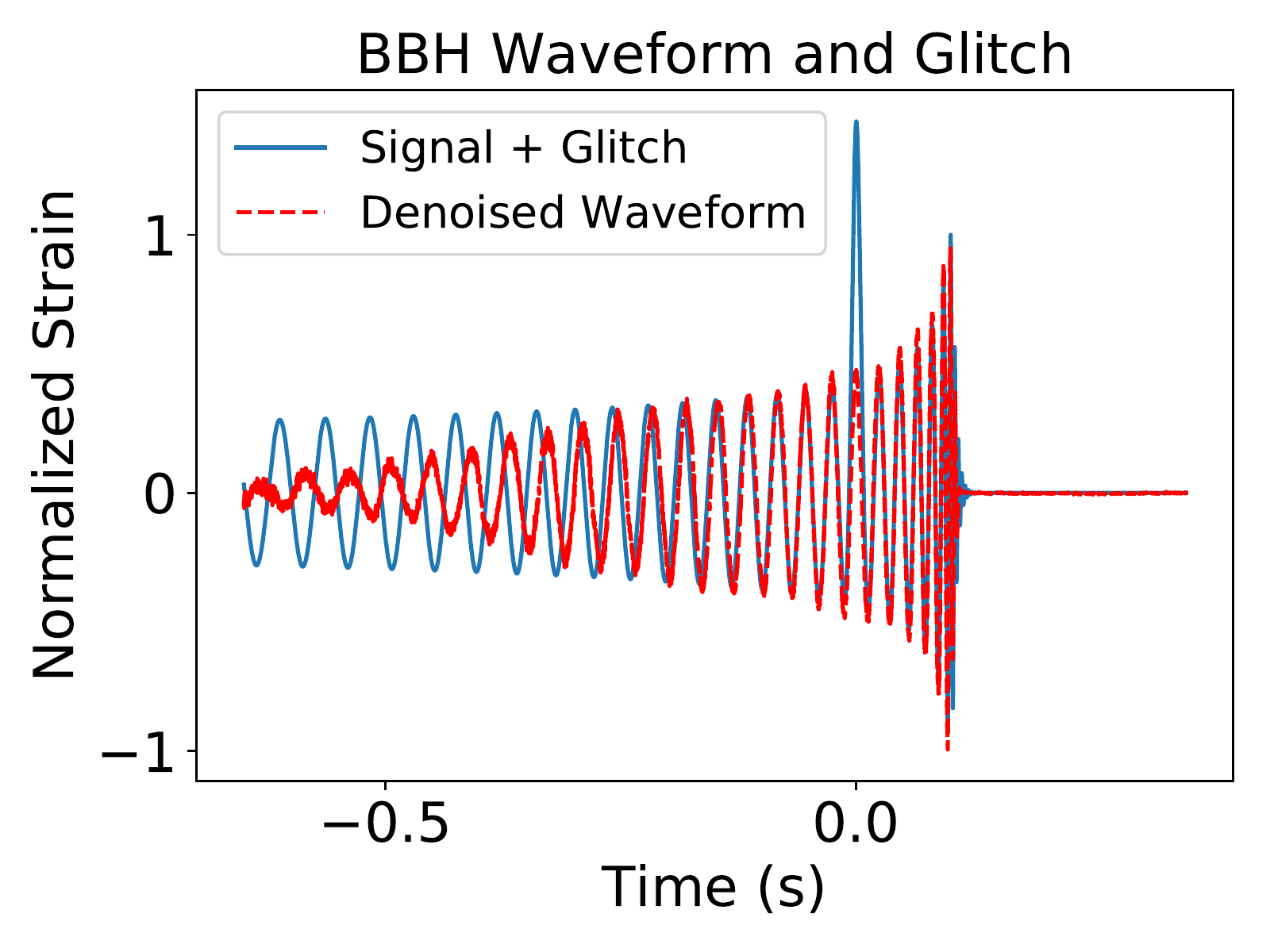}
\includegraphics[width=0.5\linewidth]{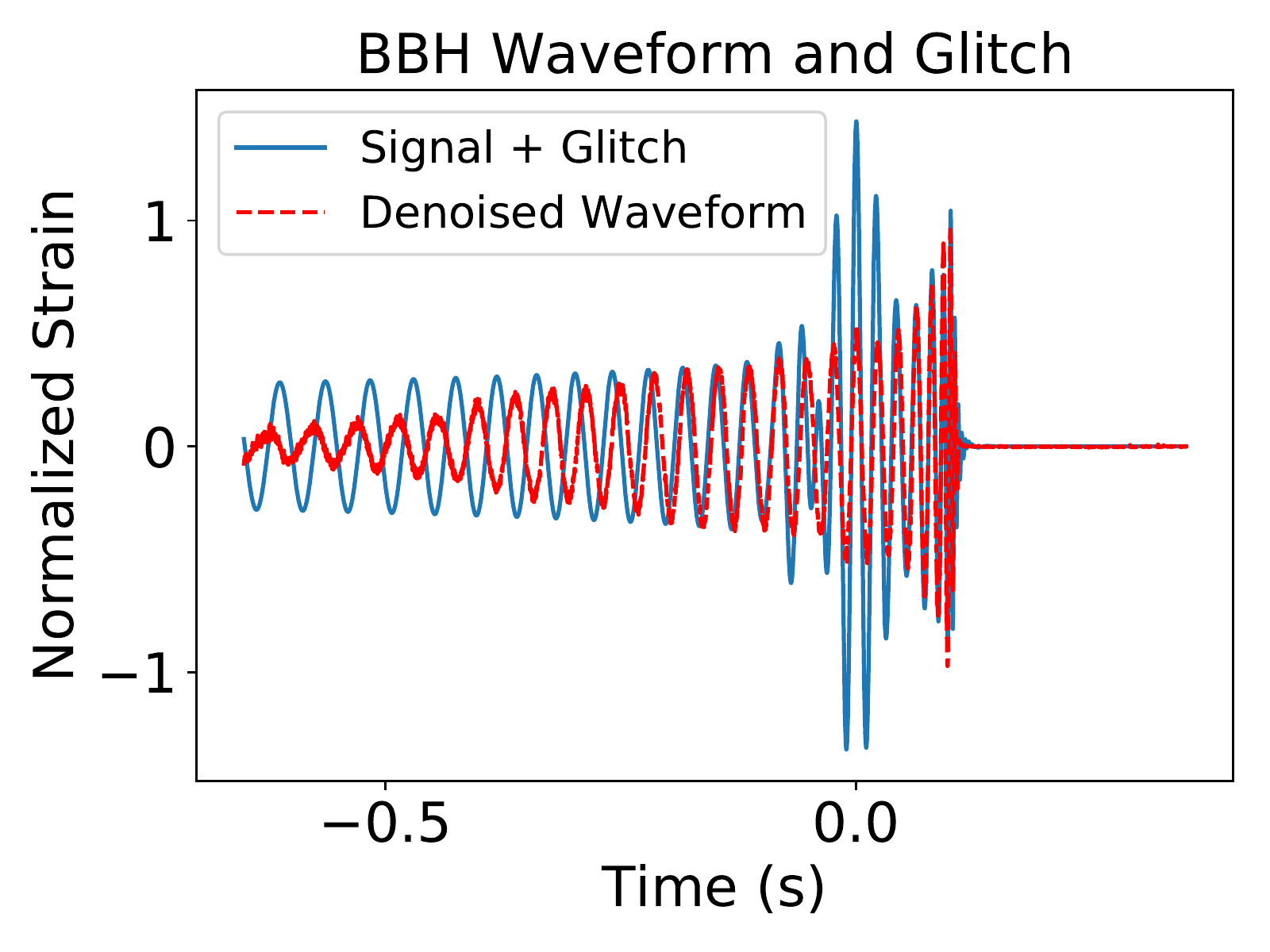}
}
\centerline{
\includegraphics[width=0.5\linewidth]{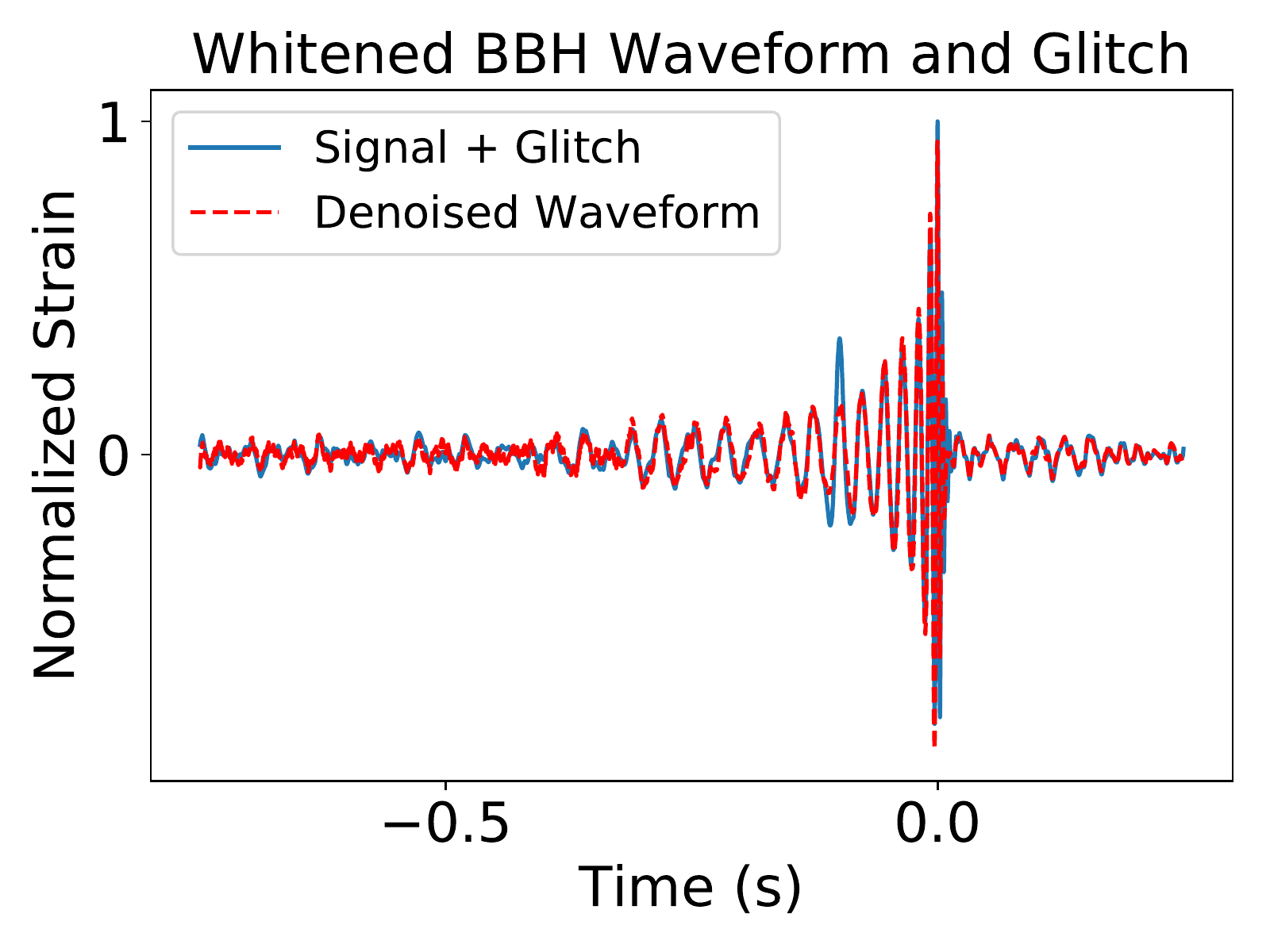}
\includegraphics[width=0.5\linewidth]{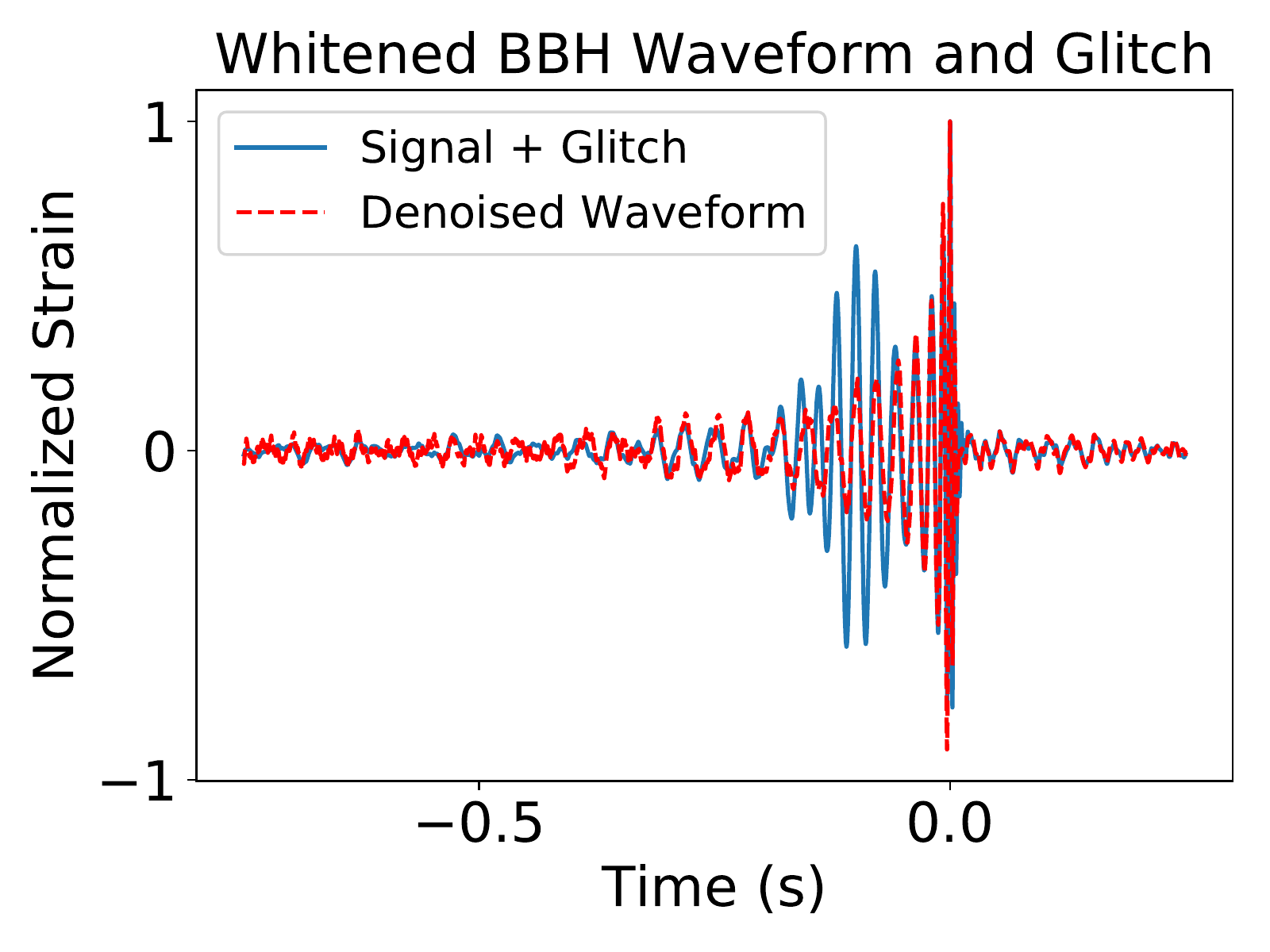}
}
\centerline{
\includegraphics[width=0.5\linewidth]{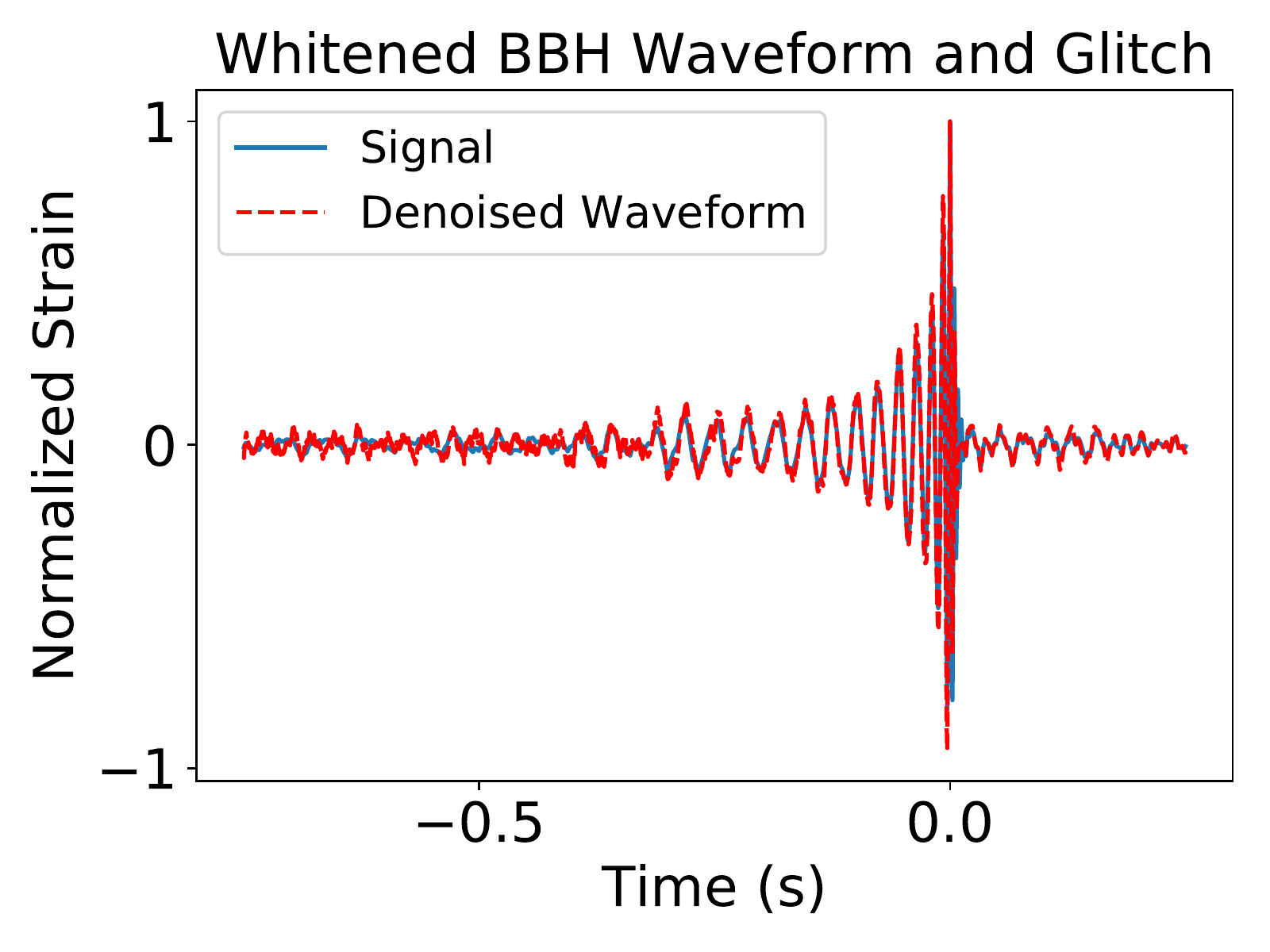}
\includegraphics[width=0.5\linewidth]{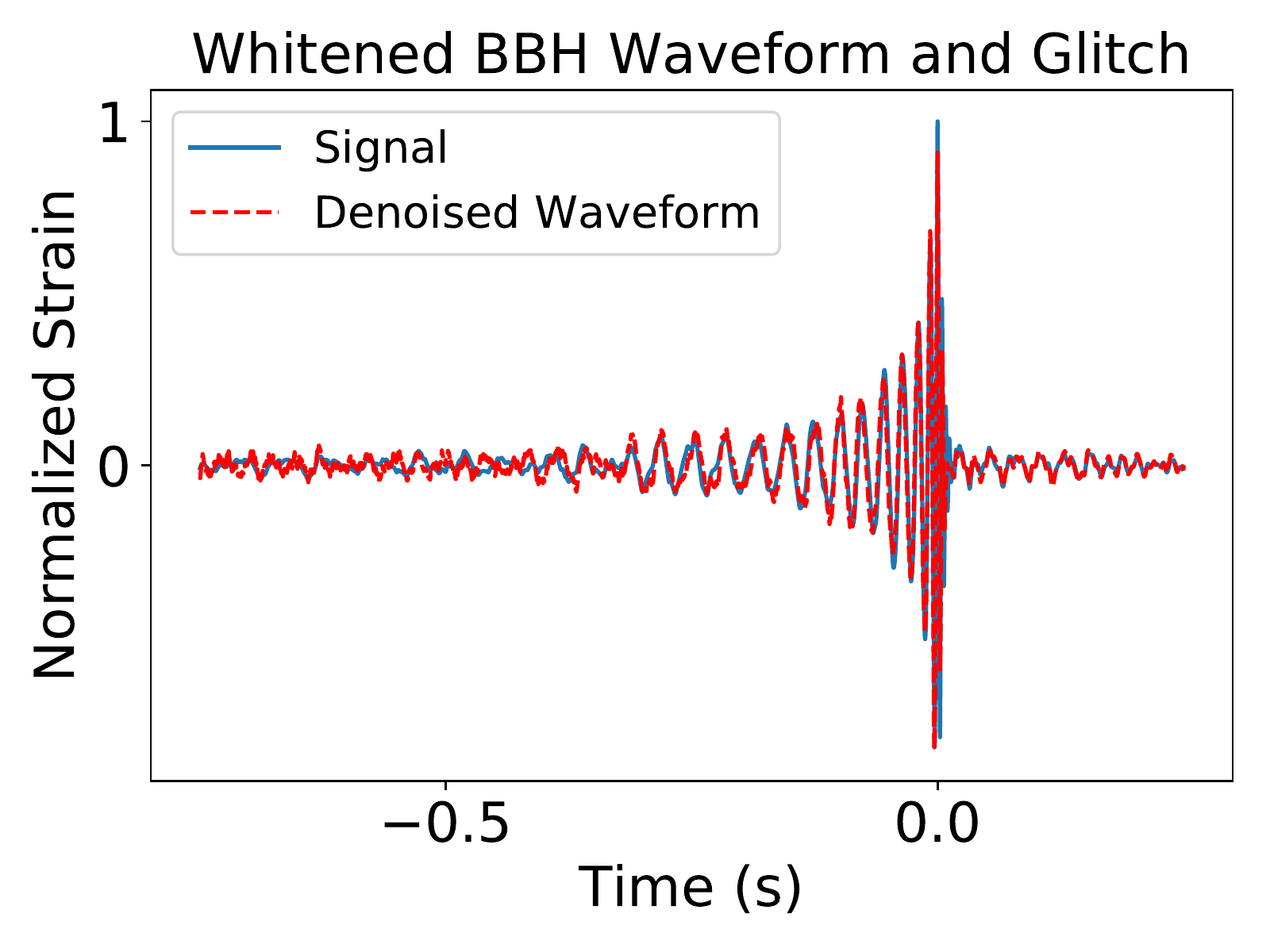}
}
\caption{These panels present signals contaminated by two types of glitches, namely, Gaussian glitch for the left panels and sine-Gaussian glitch for the right panels. The top panels present the ground-truth signals contaminated by the noise anomalies, accompanied by the output of our denoising algorithm. Notice that, as expected from an optimal denoiser, our deep learning algorithm has removed the glitches from the denoised signals. Bottom panels: whitened version of the ground-truth signals contaminated by glitches and of our denoised signals. Bottom panels: whitened version of the ground-truth signals without glitch contamination and of our denoised signals.}
\label{fig:glitches_on_waveforms}
\end{figure*}

Figure~\ref{fig:glitches_on_waveforms} presents results for a variety of studies we performed with our denoising algorithm. The top panels present ground-truth signals contaminated by two types of glitches discussed in~\cite{jade:2015CQGra,jade1:2016}, and the corresponding signals that were produced by our deep learning algorithm. For this analysis we considered non-spinning BBH mergers with total mass \(M=64.5\msun\), and mass-ratio \(q=1.24\). We injected these signals in open source LIGO data in the vicinity of the event GW150914~\cite{losc}. We notice that, as expected, the denoiser has removed both types of noise anomalies from the denoised waveform signals. The mid-panels show how the actual signals (contaminated by glitches and denoised ones) look when whitened by a noise PSD estimate. The bottom panels show the whitened version of the original signals (without glitch contamination) and the denoised signals. These studies show that our denoising algorithm has learned to tell apart noise anomalies from signals, and that it is effective at removing these from waveforms.

\begin{figure*}
\centerline{
\includegraphics[width=0.5\linewidth]{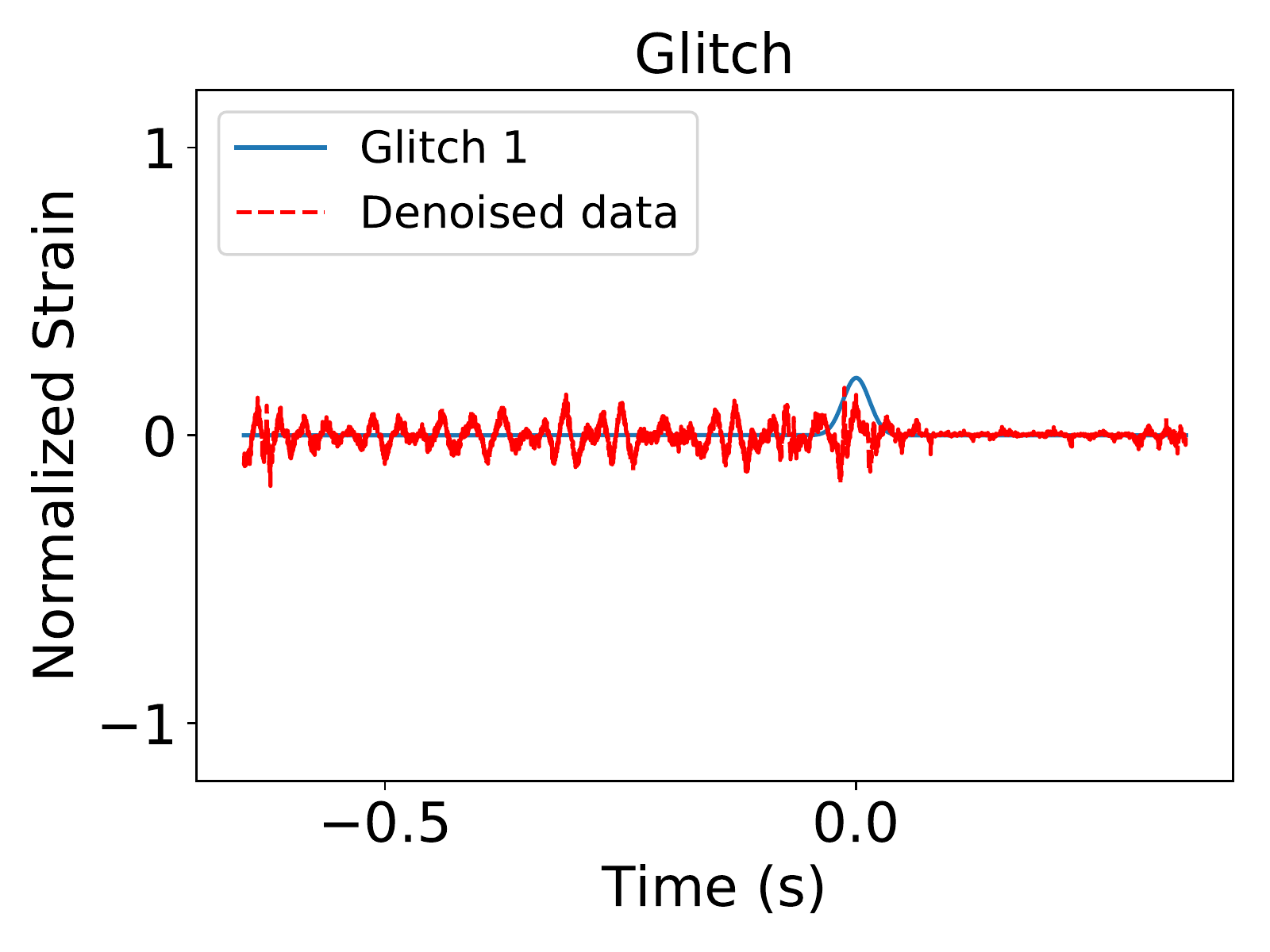}
\includegraphics[width=0.5\linewidth]{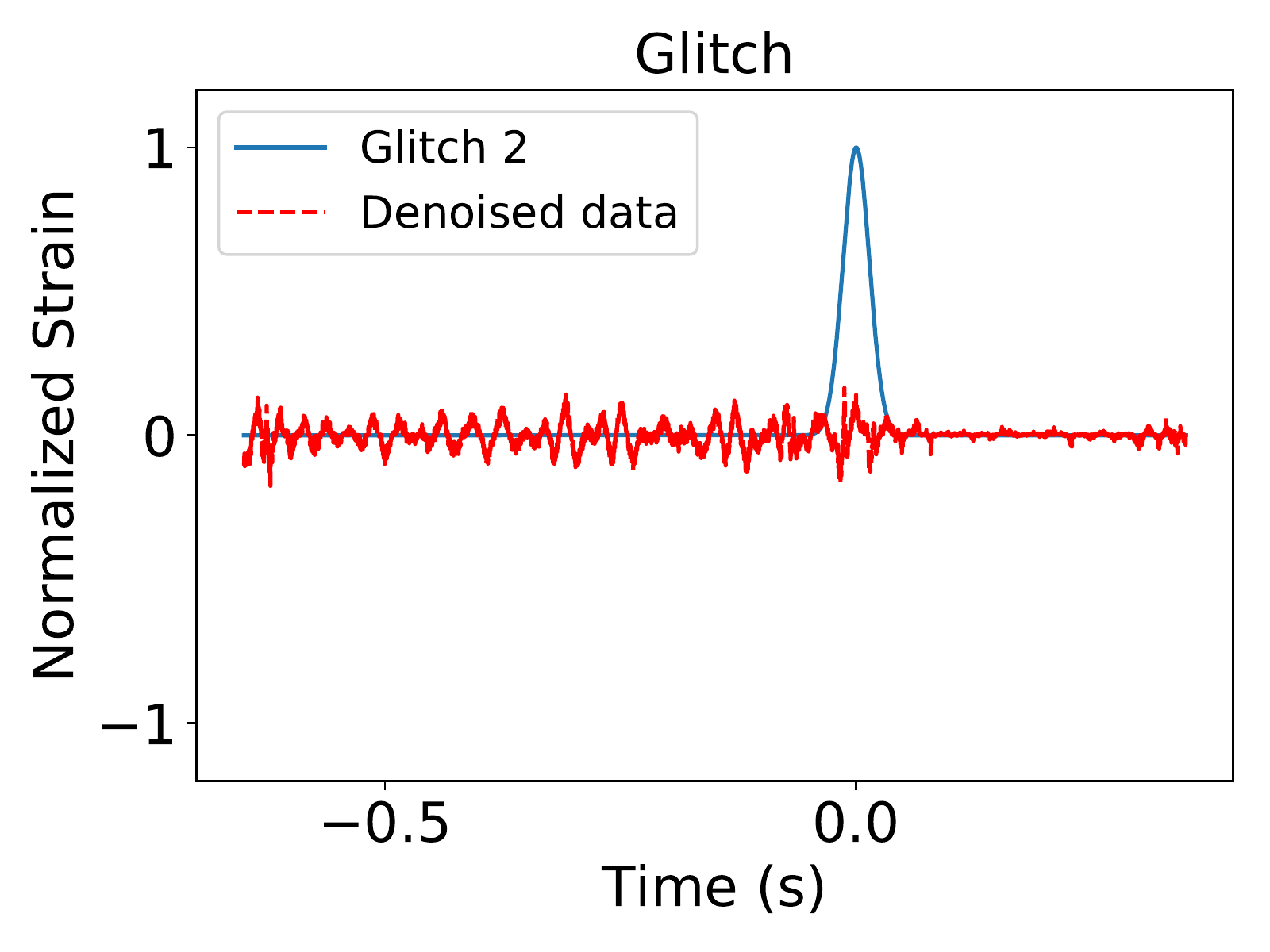}
}
\centerline{
\includegraphics[width=0.5\linewidth]{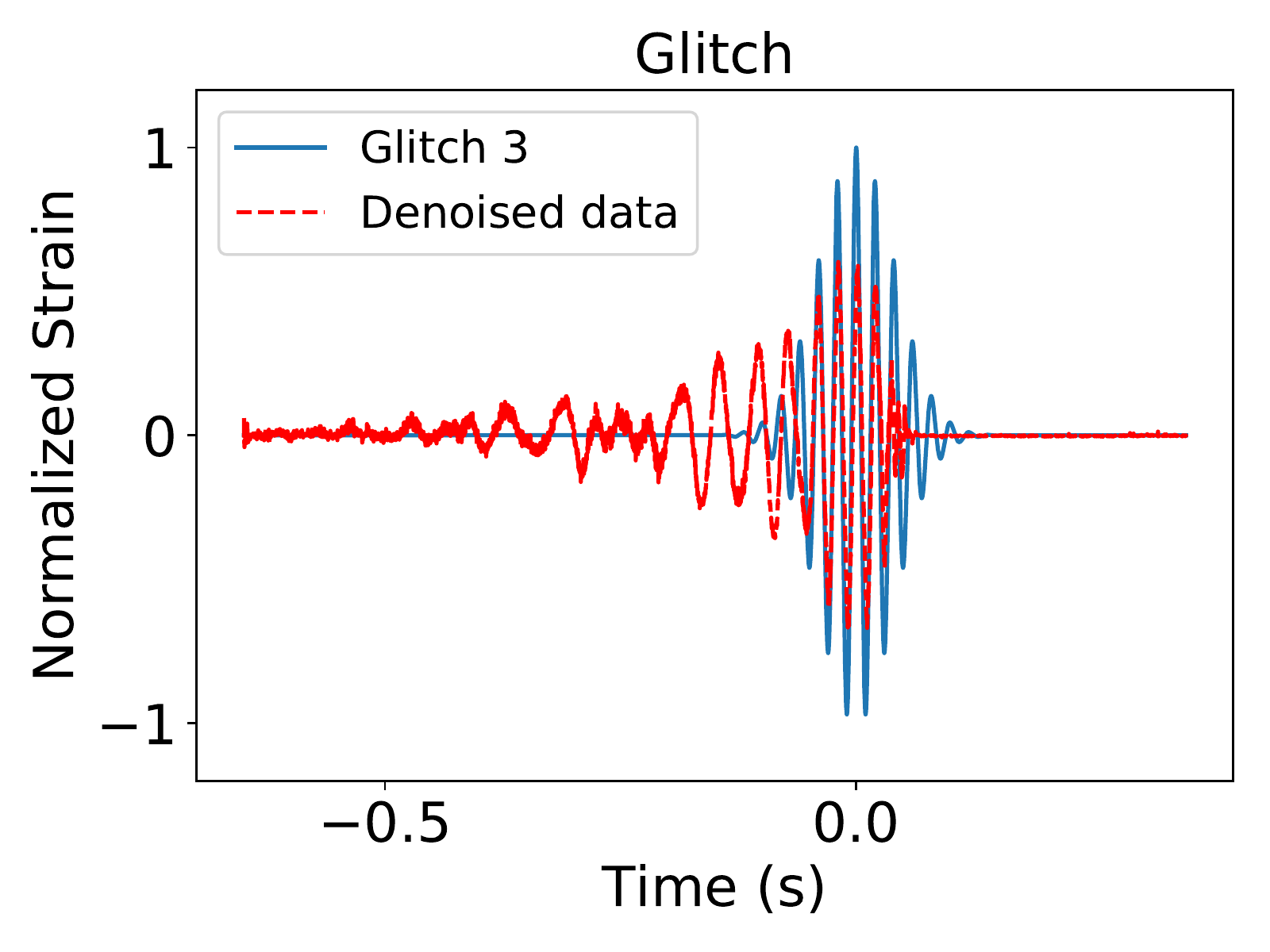}
\includegraphics[width=0.5\linewidth]{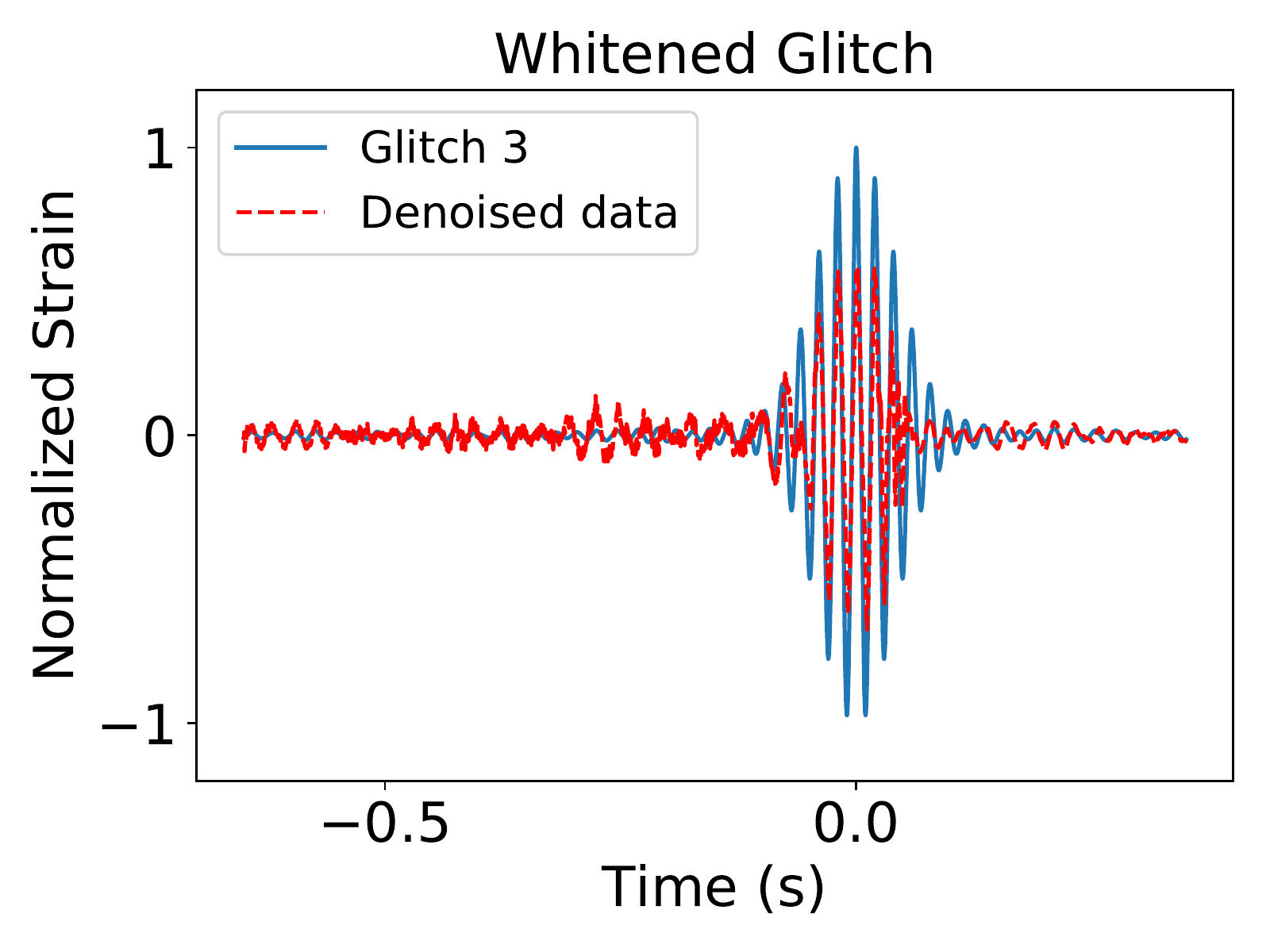}
}
\caption{The panels show the output of our denoising algorithm when it is used to process real advanced LIGO noise that contains glitches but no waveforms. Top panels: Gaussian glitches with different amplitudes are injected in real LIGO noise. The output of our deep learning algorithm is consistent with the absence of waveform signals in the data. Bottom panels: the left panels shows a sine-Gaussian glitch and its denoised version. The right panel shows the whitened version of the glitch and of the denoised glitch. This shows that structured glitches of this nature can effectively be denoised by our deep learning algorithm, reconstructing with fair-fidelity the amplitude and phase of these noise anomalies.}
\label{fig:just_glitches}
\end{figure*}

\noindent Finally, we have considered the scenario in which there is no waveform in the data, but only noise anomalies. We performed three experiments, namely, we extracted open source advanced LIGO data in the vicinity of the event GW150914~\cite{losc}, and injected two types of Gaussian glitches, as shown in the top and bottom panels of Figure~\ref{fig:just_glitches}. In the case of Gaussian glitches, we found that the output of our denoising algorithm is consistent with the expected time-series data that it would produce in the absence of waveform signals, as shown in the top panels of Figure~\ref{fig:just_glitches}. 

\begin{figure*}
\centerline{
\includegraphics[width=0.33\linewidth]{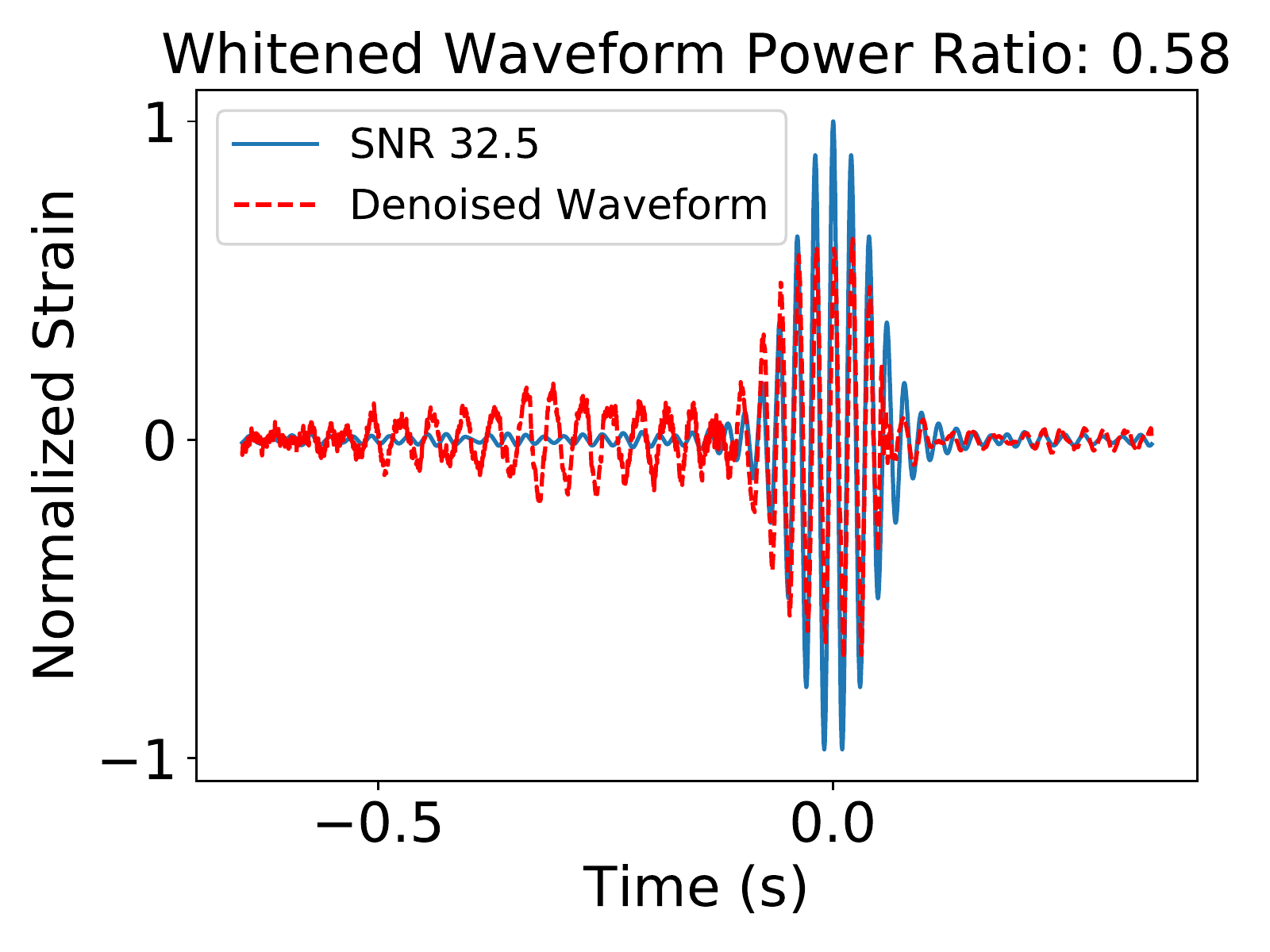}
\includegraphics[width=0.33\linewidth]{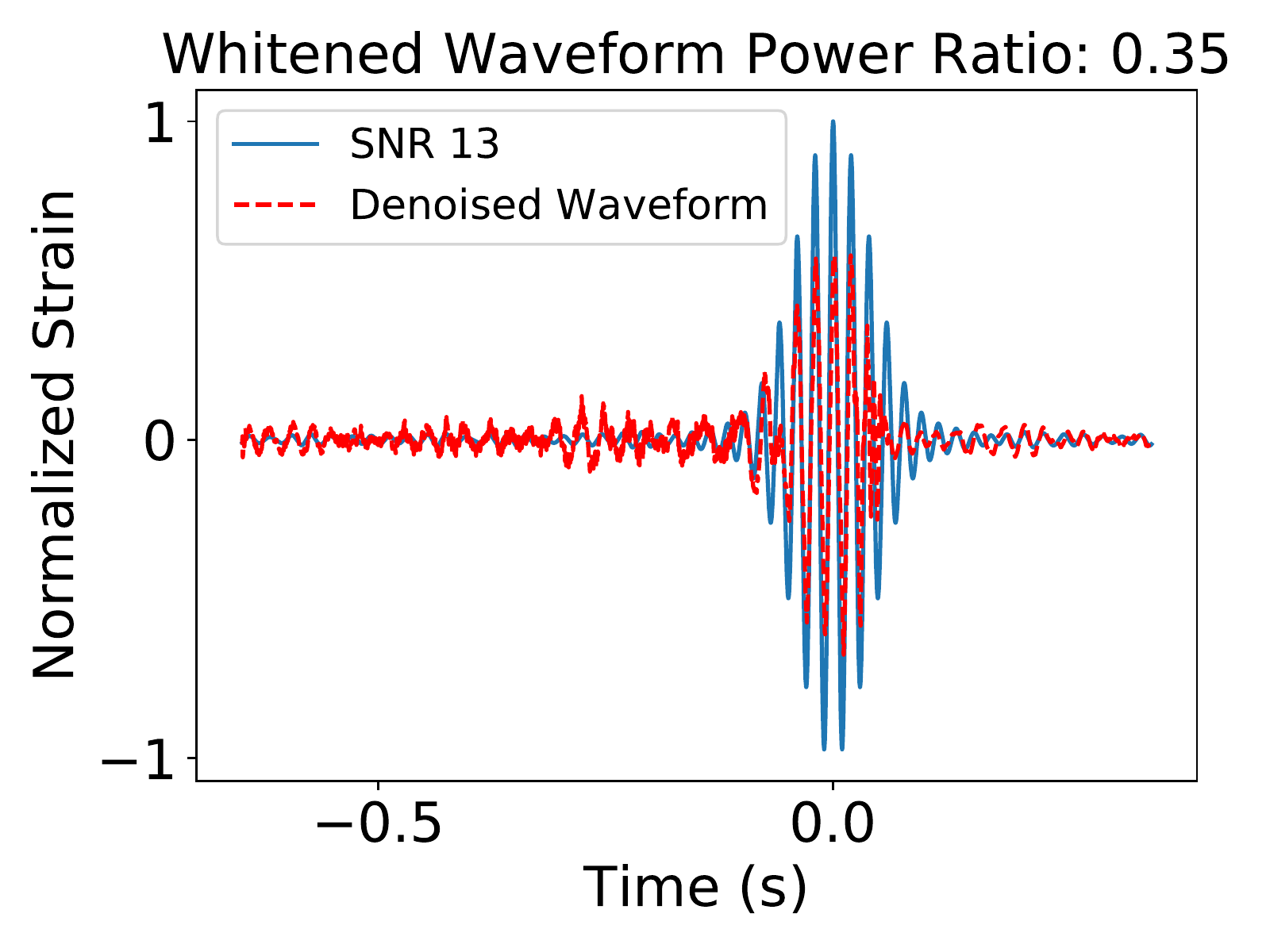}
\includegraphics[width=0.33\linewidth]{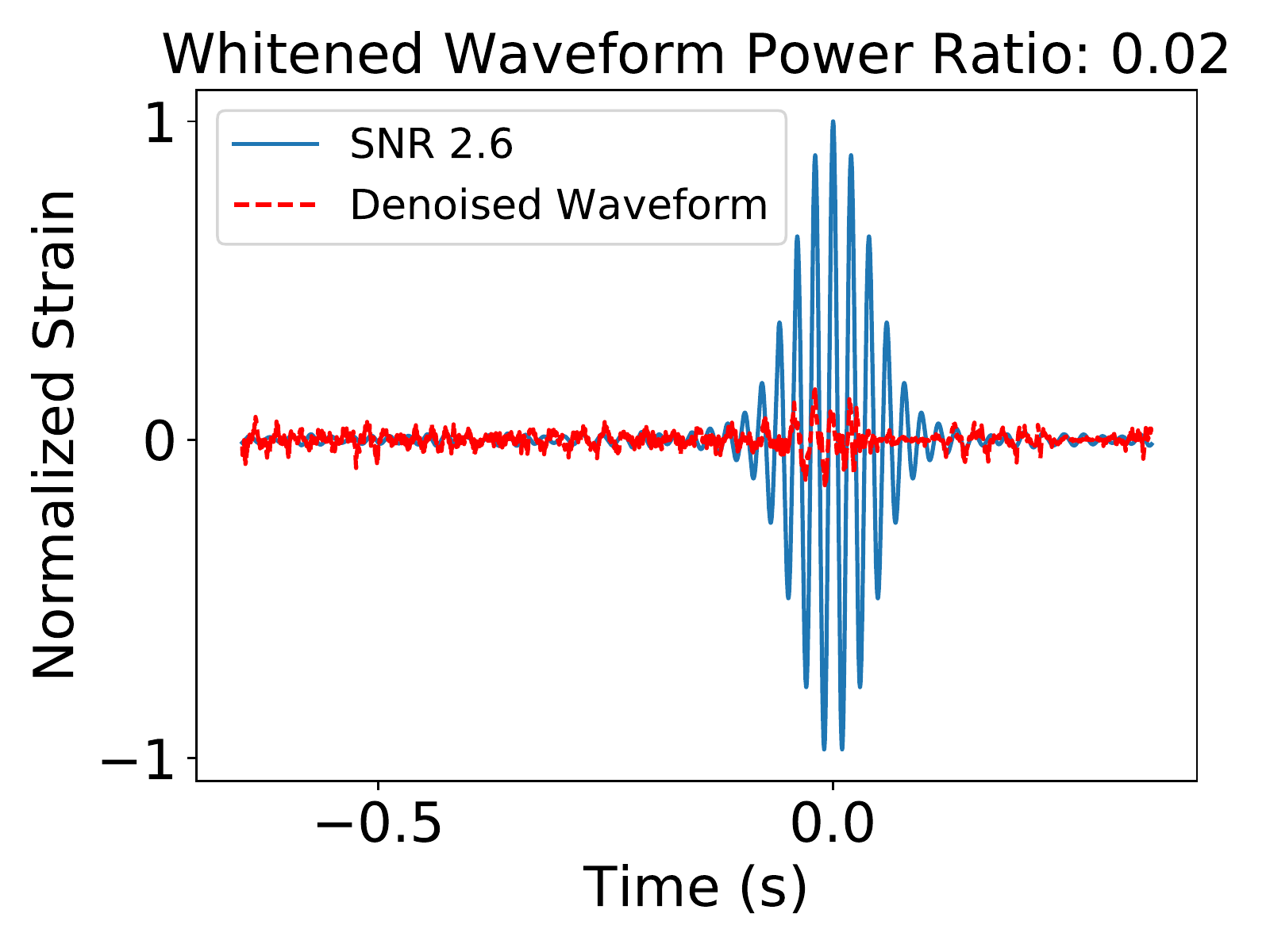}
}
\caption{From left to right, sine-Gaussian glitches with \(\textrm{SNR}=\{32.5,\,13,\,2.6\}\). The average power of the denoised glitches is, from left to right, \(P=\{58\%,\,35\%,\,2\%\}\). This indicates that while our denoiser may reconstruct the phase of structured glitches with fair fidelity, the amplitude reconstruction of these noise anomalies is suboptimal, even for loud glitches.}
\label{fig:just_glitches_power}
\end{figure*}

On the other hand, we have also considered a structured noise anomaly, namely, a sine-Gaussian glitch. As shown in the bottom panels of Figure~\ref{fig:just_glitches}, this type of glitch resembles a GW signal, and our algorithm tries to actually reconstruct it. There are key differences, however, in the output time-series signal produced by our denoiser when it reconstructs a true or simulated GW signal, and a noise anomaly that resembles a GW signal, such as a sine-Gaussian glitch. In the case of GW signals, the denoised amplitude and phase of the signal closely resemble the ground truth signal, as shown in Figure~\ref{fig:glitches_on_waveforms}. In contrast, when our denoiser is applied to GW data that only contains noise and a sine-Gaussian glitch, we find that the denoised time-series captures fairly well the phase of the ground-truth sine-Gaussian, while it poorly recovers its amplitude evolution. We have explored this latter finding in detail, and present a summary of these results in Figure~\ref{fig:just_glitches_power}. Therein we present the recovered average power, \(P\), of the signals defined as

\begin{equation}
    P = \lim_{L\to \infty}\frac{1}{2L}\int_{-L}^{L}\left|x(t)\right|^2\mathrm{d}t\,,
\end{equation}

\noindent where \(x(t)\) represent the time-series sine-Gaussian glitch. Using this relation we have computed the average power of the denoised glitches assuming three cases \(\textrm{SNR}=\{32.5,\,13,\,2.6\}\), obtaining \(P=\{58\%,\,35\%,\,2\%\}\), respectively. These results indicate that our denoiser is suboptimal to recover this type of noise anomalies.

\noindent These studies shed light on the realm of applicability of deep learning algorithms to: (i) denoise signals in realistic detection scenarios, including signals that describe a signal manifold that is distinct to the one used for training; (ii) remove glitches from waveform signals that are embedded in real advanced LIGO noise; (iii) tell apart signals from glitches. We have also identified specific areas of improvement, including the development of a denoiser that is trained with spinning BBH mergers. This tools will be presented in future work.
%%%%%%%%%%%%%%%%%%%%%%%%%%%%%%%%%%%%%%%%%%%%%%%%%%%%%%
%%%%%%%%%%%%%%%%%%%%%%%%%%%%%%%%%%%%%%%%%%%%%%%%%%%%%%
\section{Conclusions}
\label{end}

We have designed deep learning algorithms to denoise GW signals embedded in simulated Gaussian noise, and in realistic detection scenarios, using non-Gaussian and non-stationary LIGO noise. In the former case, we have demonstrated that the overlap between the output time-series signal of our denoising algorithm, and the ground truth signals is \({\cal{O}}\geq0.97\) across the BBH parameter space \(m_{\{1,2\}}\in[5\msun,\,75\msun]\) for noisy signals with \(\textrm{SNR}\geq12\). 

When applied to a variety of GW signals that describe spinning BBH mergers detected by advanced LIGO and Virgo, we have shown that the overlap between the output of our deep learning algorithm and the NR templates that optimally describe these events has overlaps  \({\cal{O}}\geq0.99\). 

We have also used GW170608 to demonstrate that the quality of the denoised GW is determined by the loudness of the signal and the sensitivity of the detector. We also showed that our deep learning algorithm can generalize to new types of sources, denoising spin-precessing BBH mergers. In this region of parameter space, the overlap between the ground-truth signals and the output of our denoiser is \({\cal{O}}\geq0.97\). Finally, we showed that deep learning can remove noise anomalies from BBH mergers.

This work, combined with other successful efforts using machine learning to denoise time-series signals embedded in simulated or non-Gaussian and non-stationary noise, suggest that instead of designing sophisticated schemes to model the statistical properties of noise, one may use deep learning algorithms to \textit{learn} the true properties of noise, and use this knowledge to carry out controlled experiments in which modeled signals are embedded and subsequently extracted from realistic noise datasets. 

Furthermore, for GW signals with moderate to GW150914-type SNRs, we can readily use this denoising algorithms to process data segments where true GW signals are marginally detected as a result of contamination from noise anomalies, or to assess whether single-interferometer observations actually have signals in other detectors which have been obscured by noise anomalies. Furthermore, denoised signals may also be used as input data for other deep learning algorithms that may provide point-parameter estimation results of Bayesian deep learning parameter estimation analyses~\cite{Shen:2019vep,bambiann:2015PhRvD,mau:2018aM}. 

In future work, we will design neural network models to denoise longer GW signals. We will also explore the applicability of these methodologies to extract and denoise new classes of GWs from other realistic noise datasets, targeting in particular potential GW sources that may be observed with pulsar timing arrays~\cite{11yrdataset:2018,lam:2018aL,lam:2018aL}. 

%%%%%%%%%%%%%%%%%%%%%%%%%%%%%%%%%%%%%%%%%%%%%
%%%%%%%%%%%%%%%%%%%%%%%%%%%%%%%%%%%%%%%%%%%%%
\section{Acknowledgments}
\label{ack}

EAH and WW gratefully acknowledge National Science Foundation (NSF) awards OAC-1931561 and OAC-1934757.

This research is part of the Blue Waters sustained-petascale computing project, 
which is supported by the NSF (awards OCI-0725070 and ACI-1238993) 
and the State of Illinois. Blue Waters is a joint effort of the University of Illinois at 
Urbana-Champaign and its National Center for Supercomputing Applications. 
We acknowledge support from the NCSA. We thank the 
\href{http://gravity.ncsa.illinois.edu}{NCSA Gravity Group} for useful feedback, 
and Vlad Kindratenko for granting us access to state-of-the-art GPUs and HPC 
resources at the Innovative Systems Lab at NCSA. We are grateful to NVIDIA 
for donating several Tesla P100 and V100 GPUs that we used for our analysis, 
and compute resources at the Illinois Campus Cluster Program~\cite{CCUUIUC}.

This work utilizes resources supported by the National Science Foundation's Major Research Instrumentation program, grant \#1725729, as well as the University of Illinois at Urbana-Champaign.

This work used the Extreme Science and Engineering Discovery Environment (XSEDE), 
which is supported by National Science Foundation grant number ACI-1548562. 
Specifically, it used the Bridges system, which is supported by NSF award number 
ACI-1445606, and TG-PHY160053, at the Pittsburgh Supercomputing Center (PSC).

This research used resources of the Argonne Leadership Computing Facility, which is a DOE Office of Science User Facility supported under Contract DE-AC02-06CH11357.

\section*{References}

\bibliography{ref_one,ref_two,nr_refs}

\appendix

\section{Denoising signals in simulated Gaussian noise}
\label{app1}

\begin{figure}[h]
    \centering
    \includegraphics[width=0.85\linewidth]{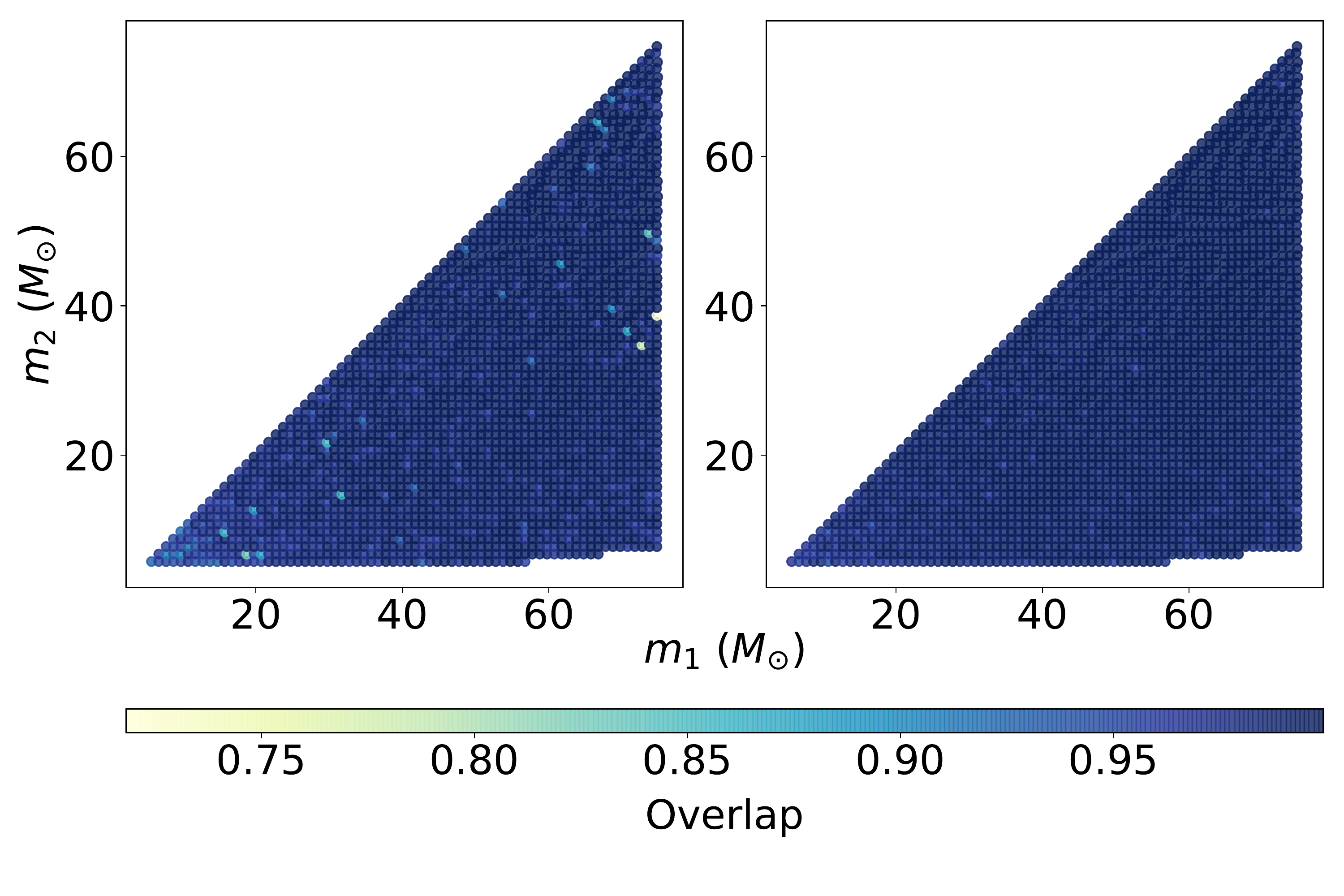}
    \caption{Normalized overlap between the output signal of our denoising algorithm (Model \RNum{2}), and the corresponding clean signals using BBH populations with matched-filtering \(\textrm{SNR}=9\) (left panel) and matched-filtering \(\textrm{SNR}=12\) (right panel).}
    \label{fig:gaussian_256}
\end{figure}

\noindent Figure~\ref{fig:gaussian_256} presents the reconstruction accuracy of noisy BBH GW signals embedded in Gaussian noise. These results were obtained using Model \RNum{2}, described in Section~\ref{m2}. We notice that our denoising algorithm produces time-series signals whose properties reproduce the true signals with accuracies \(\geq97\%\) for noisy signals with \(\textrm{SNR}\geq12\).

\end{document}